\documentclass[
superscriptaddress,
tightenlines,
nofootinbib,
twocolumn,
 amsmath,amssymb,
 aps,
prd,
]{revtex4-1}

\usepackage{amsmath}
\usepackage{amssymb}
\usepackage{bm}
\usepackage{color,graphicx}
\usepackage{colortbl}
\usepackage{slashed}
\usepackage{comment}
\usepackage{mathtools}
\usepackage{booktabs}
\usepackage{xcolor}

\newcounter{comment}

\usepackage{enumerate}
\usepackage{amssymb}

\begin{document}

\title{VAIM - CFF: A variational autoencoder inverse mapper solution to Compton form factor extraction from deeply virtual exclusive reactions \begin{center} \vspace{0.5cm}
\includegraphics[scale=0.35]{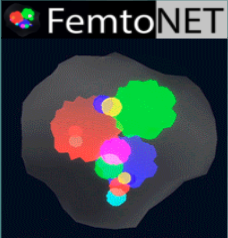}\end{center}}

\author{Manal Almaeen} 
\email{mhmaeen@ju.edu.sa}
\affiliation{Department of Software Engineering, College of Computer and Information Sciences, Jouf University, Sakaka 72341, Aljouf, Saudi Arabia.}

\author{Tareq Alghamdi} 
\email{talgh001@odu.edu}
\affiliation{Department of Computer Science, Old Dominion University, Norfolk, VA 23529, USA.}
\affiliation{Department of Computer Science, College of Computer Science and Information Technology, Al-Baha University, Al-Baha, Alaqiq 65779, Saudi Arabia}

\author{Brandon Kriesten} 
\email{bkriesten@anl.gov}
\affiliation{High Energy Physics Division, Argonne National Laboratory, Lemont, IL 60439}

\author{Douglas Adams}  
\email{yax6jr@virginia.edu}
\affiliation{Department of Physics, University of Virginia, Charlottesville, VA 22904, USA.}

\author{Yaohang Li} 
\email{yaohang@cs.odu.edu}
\affiliation{Department of Computer Science, Old Dominion University, Norfolk, VA 23529, USA.}

\author{Huey-Wen Lin} 
\email{hwlin@pa.msu.edu}
\affiliation{Department of Physics, Michigan State University, East Lansing, MI 48824, USA.}

\author{Simonetta Liuti} 
\email{sl4y@virginia.edu}
\affiliation{Department of Physics, University of Virginia, Charlottesville, VA 22904, USA.}



\begin{abstract}
We develop a new methodology for extracting Compton form factors (CFFs) in from deeply virtual exclusive reactions such as the unpolarized DVCS cross section using a specialized inverse problem solver, a variational autoencoder inverse mapper (VAIM). The VAIM-CFF framework not only allows us access to a fitted solution set possibly containing multiple solutions in the extraction of all 8 CFFs from a single cross section measurement, but also accesses the lost information contained in the forward mapping from CFFs to cross section. We investigate various assumptions and their effects on the predicted CFFs such as cross section organization, number of extracted CFFs, use of uncertainty quantification technique, and inclusion of prior physics information. We then use dimensionality reduction techniques such as principal component analysis to visualize the missing physics information tracked in the latent space of the VAIM framework. Through re-framing the extraction of CFFs as an inverse problem, we gain access to fundamental properties of the problem not comprehensible in standard fitting methodologies: exploring the limits of the information encoded in deeply virtual exclusive experiments. 
 \end{abstract}

\maketitle

\allowdisplaybreaks
\newpage
\section{Introduction}
\label{sec:intro}


Deeply virtual exclusive scattering processes, such as deeply virtual Compton scattering, have been identified as the ``golden channel" for the extraction of information on partonic 3D dynamics in the nucleon. In the context of QCD factorization theorems  \cite{Collins:1996fb, Collins:1998be, Muller:1994ses}, this scattering can be parameterized by non-local, off-forward matrix elements, the generalized parton distributions (GPDs)~\cite{Ji:1996ek,Muller:1994ses,Radyushkin:1997ki}, which contain new information on hadronic structure beyond the longitudinal distributions probed in inclusive scattering such as transverse spatial densities, angular momentum, and pressure/shear forces (for  reviews  see Refs.~\cite{Diehl:2003ny,Belitsky:2005qn,Kumericki:2016ehc}).

In DVCS and similar exclusive reactions the virtual photon from the scattered lepton cannot directly probe the full $x$ dependence of the non-local matrix elements, or GPDs ($x$ being the momentum fraction taken by the probed parton). The scattering amplitude for the DVCS process is, in fact, written in terms of non-perturbative structures, the Compton form factors (CFFs), which are convolutions in $x$ of the GPDs with non-trivial QCD Wilson coefficient functions~\cite{Ji:1996nm}.  This is juxtaposed to inclusive scattering cross sections where the non-perturbative matrix elements, the parton distribution functions (PDFs), enter directly at the cross section, level. This added layer of complexity makes the analysis of DVCS processes sensitive to the extraction method, and requires the development of new techniques to disentangle these convolutions. It then becomes pertinent to understand how much information regarding nuclear structure is retained through these layers of abstraction and therefore accessible from experiment. 
A comprehensive treatment of the formalism for DVCS was given in Ref.~\cite{Kriesten:2019jep}.

To address these problems in a scientifically rigorous way, the FemtoNet group leveraged physics informed neural network architectures to extract the maximum amount of information from experimental observables as possible. 
\footnote{The Femtonet collaboration, with an augmented membership. is presently transitioning into the EXCLusives with Artificial Intelligence and Machine learning (EXCLAIM) collaboration.}
Our neural network pipeline is:
\begin{enumerate}
    \item Cross section analysis and modelling / generalization
    \item CFF fitting from observables and inverse problems with understandable errors
    \item GPD extractions and parameterizations
    \item Physical properties of interest: angular momentum, pressure distributions, and spatial densities
\end{enumerate}
Critical to this pipeline is a set of benchmarks to standardize various extraction techniques. These benchmarks were first written down in Ref.~\cite{Almaeen2021}, where it was suggested that an analysis of the unpolarized cross section is a critical benchmark process. Understanding the unpolarized cross section is vital for any other DVCS observable such as beam/target polarization asymmetries. The extraction of CFFs from the unpolarized cross section has been the subject of many QCD analyses for the past two decades (see  Ref.~\cite{Kumericki:2016ehc} for a review).

In this manuscript we re-frame the analysis of the unpolarized cross section to focus on the quantity and quality of information contained in single experimental observables. To accomplish this, we treat the extraction of CFFs from a single polarization observable as an inverse problem from which there can be infinite number of solutions. Since the unpolarized cross section is parameterized by eight unknowns, the extraction of all CFFs is an ill-posed inverse problem.

We utilize a variational autoencoder inverse mapper (VAIM)~\cite{9534012} to frame the inverse problem of CFF extractions and determine CFF solutions with propagated experimental errors.  
The main advantage of using an autoencoder for the physics problem at hand is in that through the process of first encoding the input data into a reduced dimensionality subspace, and subsequently  reconstructing/decoding them, the autoencoder focuses on and retains the essential, valuable information buried in the original data set. 

It has already been shown that the VAIM framework is a crucial tool in understanding inverse problems in high energy physics, including the extraction of the PDF parameters from fits to deep inelastic scattering (DIS) data~\cite{Almaeen:2022ifg}. Through the use of data augmentation we can simultaneously utilize the physics information carried in the experimental errors while also increasing the size of our dataset, therefore reducing systematic neural network uncertainties. 

This paper is structured as follows: in Sec.~\ref{sec:background} we give a theoretical motivation for the physics, machine learning, uncertainty quantification, and the data used. In Sec.~\ref{sec:vaim_background} we motivate the VAIM architecture used to solve the inverse problem of CFF extraction. Section~\ref{sec:results} is dedicated to the results of our analysis. We draw our conclusions in Sec.~\ref{sec:conclusion}.

\section{Background}
\label{sec:background}
\subsection{Theoretical Framework}
\label{sec:theory}

Deeply virtual exclusive reactions such as DVCS are a special class of scattering processes in which the initial proton fragments into an intermediate quantum state which propagates before emitting a real photon and a proton. The non-perturbative structures that describe this process are called GPDs and they encode not only longitudinal momentum information of the partons, but also spatial information through the Fourier transform in the momentum transfer between the initial and final states (the physics of GPDs is reviewed in Refs.~\cite{Diehl:2003ny,Belitsky:2005qn}). 

A key observation was made in Ref.~\cite{Ji:1996ek} that the local matrix elements of the energy momentum tensor (EMT), describing the gravitational interactions of the nucleon are connected to the Mellin moments of the GPDs through the operator product expansion. Therefore the scattering of a massless spin-2 boson off of the nucleon can be re-interpreted in terms of the interaction of two photons with the nucleon. Specifically, 
one obtains the following sum rule \cite{Ji:1996ek},  
\begin{eqnarray}
J^{q,g} &=& \frac{1}{2}\Big(A_{2,0}^{q,g}(0) + B_{2,0}^{q,g}(0) \Big)
\end{eqnarray}
where the superscripts $q$ and $g$ refer to the quark and gluon, respectively, and the connection between form factors and Mellin moments of the GPDs, $H^{q,g}$ and $E^{q,g}$, is given by,
\begin{eqnarray}
\int_{-1}^{1}dx x H^{q,g}(x,\xi,t) &=& A_{2,0}^{q,g}(t) + 4\xi^{2}C_{2}^{q,g}(t) \\
\int_{-1}^{1}dx x E^{q,g}(x,\xi,t) &=& B_{2,0}^{q,g}(t) - 4\xi^{2}C_{2}^{q,g}(t).
\end{eqnarray}
\noindent The GPDs appear in the DVCS cross section convoluted through integrals in $x$ with perturbatively calculable Wilson coefficient functions, $\mathcal{C}^{\pm}$, called the Compton form factors (CFFs), $\mathcal{H}^{q}$ and $\mathcal{E}^{q}$. Two extra CFFs, $\widetilde{\mathcal{H}}^{q}$, and $\widetilde{\mathcal{E}}^{q}$, also appear in the cross section in the axial-vector sector. The separation of these CFFs into real and imaginary parts is described in Appendix~\ref{sec:app_cff}. 

The total DVCS cross section is dependent on kinematic variables $(x_{Bj},t,Q^{2},E_{b},\phi)$ where,
\begin{itemize}
    \item $x_{Bj}=Q^{2}/2(pq)$, Bjorken-$x$, is the fraction of longitudinal momentum carried by the struck parton in the limit $Q^{2},(pq) \rightarrow \infty$ and $x_{Bj}$ stays constant. 
    \item $Q^{2} = -q^{2}$ is the invariant mass of the exchanged virtual photon.
    \item $t = (p-p')^{2}$ is the square of the 4-momentum transfer between initial and final proton states.
    \item $E_{b}$ is the electron beam energy.
    \item $\phi$ is the azimuthal angle between lepton and hadron scattering planes.
\end{itemize}
A background process to DVCS, Bethe-Heitler (BH) scattering~\cite{osti_4064532, Gehrmann:1997qh}, is also present where the final photon is emitted from either the initial or final electrons, and the hadronic matrix elements are proportional to the proton elastic form factors. The cross section for the $e p \rightarrow e' p' \gamma'$ cross section is therefore given by,
\begin{eqnarray}
\sigma &=& \sigma_{o}|\mathcal{T}_{BH} + \mathcal{T}_{DVCS}|^{2}
\end{eqnarray}
where $\sigma_{o}$ is the elementary scattering cross section.
By squaring the sum of the amplitudes, there exists an interference cross section which is linearly dependent on CFFs with amplified kinematic dependence from the BH process.
\begin{eqnarray}
\sigma &=& \sigma_{BH} + \sigma_{DVCS} + \sigma_{\mathcal{I}}
\end{eqnarray}
where the interference term is defined as,
\begin{eqnarray}
\sigma_{\mathcal{I}} &=& 2\Im m \Big(\mathcal{T}_{BH}^{*} \mathcal{T}_{DVCS} \Big) .
\end{eqnarray}
The total kinematic dependence of these cross sections have been worked out many times over (for a review of the DVCS formalism and comparison of other works in the literature, see Refs.~\cite{Kriesten:2019jep,Kriesten:2020wcx} and references therein). We restrict ourselves to the unpolarized cross section. 
The latter is parametrized in terms of four complex CFFs (${\cal H}, {\cal E},  \widetilde{\cal H} and \widetilde{\cal E} )$ which describe all possible quark polarization configurations inside the proton, and of the Dirac, $F_1$, and Pauli, $F_2$, proton elastic form factors. 
The CFFs enter the cross section multiplied by coefficients that depend on various kinematic variables: $(x_{Bj}, t, Q^{2}, E_{b}, \phi)$, where $Q^2$ is the four-momentum transfer between the initial and final electrons; $x_{Bj}= Q^2/(2M \nu)$, $\nu$ being the the energy transfer, and $M$ the proton mass; $t$ is the momentum transfer between the initial and final proton; $E_b$ the electron beam energy and $\phi$ is the angle between the electrons plane and the scattering plane defined by the outgoing photon and proton. Since we will be using a subset of the world data from Jefferson Laboratory, we present the cross section for a fixed target experiment. As discussed in Refs.~\cite{Kriesten:2019jep,Kriesten:2020wcx}, the formalism can be easily extended to the collider setting, where $E_b$ is replaced by the electron-proton center of mass energy squared, $s$.   
\begin{widetext}
\begin{eqnarray}
\label{eq:BHxsec}
\sigma_{BH}(x_{Bj},t,Q^{2},E_{b},\phi) &=& \frac{\Gamma}{t}\Big[A_{UU}^{BH}\big(F_1^2 + \tau F_2^2 \big) + B_{UU}^{BH} \tau G_M^2(t) \Big]\\
\label{eq:Ixsec}
\sigma_{\mathcal{I}}(x_{Bj}, t, Q^{2}, E_{b}, \phi) &=& \frac{\Gamma}{Q^{2}t} \Big[ A_{\mathcal{I}}(x_{Bj},t,Q^{2},E_{b},\phi)\Big(F_{1}(t) \, {\bf \Re e \mathcal{H}} + \tau F_{2}(t) \, {\bf \Re e \, \mathcal{E}} \Big) \nonumber \\
&+& B_{\mathcal{I}}(x_{Bj},t,Q^{2},E_{b},\phi)G_{M}(t) \, \Big( {\bf \Re e \mathcal{H}}  + {\bf \Re e \mathcal{E}} \Big)
+ C_{\mathcal{I}}(x_{Bj},t,Q^{2},E_{b},\phi)G_{M}(t) \, {\bf \Re e \widetilde{\mathcal{H}}} \Big] \\
\label{eq:DVCSxsec}
\sigma_{DVCS}(x_{Bj},t,Q^{2},E_{b},\phi) &=&  \frac{\Gamma}{Q^{2}}\frac{2}{1-\epsilon} \Big[ (1-\xi^2)\Big[  {\bf ( \Re e {\cal H})^2} +  {(\bf \Im m {\cal H})^2} + {\bf (\Re e \widetilde{\cal H})^2} + {\bf (\Im m \widetilde{\cal H})^2} \Big] \nonumber \\
&+& \displaystyle\frac{t_o-t}{4M^2}  
\left[ {\bf (\Re e{\cal E})^2} + {\bf (\Im m{\cal E})^2}  + 
 \, {\bf \xi^2  (\Re e\widetilde{\cal E})^2} +   \, {\bf \xi^2 (\Im m\widetilde{\cal E})^2 } \right] \nonumber \\
&-&  2\xi^{2} \,\left( {\bf \Re e {\cal H} } \, {\bf \Re e {\cal E } } + {\bf \Im m {\cal H} }{\bf \Im m{\cal E } } +  {\bf \Re e \widetilde{\cal H}} \,  {\bf \Re e\widetilde{\cal E }} + {\bf \Im m \widetilde{\cal H} } {\bf \Im m \widetilde{\cal E }}    \right) \Big] .
\end{eqnarray}
\end{widetext}
The unknown quantities to extract from the cross section are the CFFs (highlighted in bold in the formulae above). All of the kinematic variables as well as the elastic form factors, $F_1$ and $F_2$  are entered as input variables.
Some further comments on the cross section are:
\begin{itemize}
    \item The BH cross section $\sigma_{BH}$ is not dependent on CFFs, and the kinematic coefficients, $A_{UU}^{BH}$ and $B_{UU}^{BH}$ are exactly calculable (Ref.\cite{Kriesten:2019jep}). 
    Therefore, to the extraction of CFFs (and within the errors of any extraction method so far), this term is simply a known constant.
    \item The BH/DVCS interference cross section is parameterized by three CFFs $\Re e \mathcal{H}$, $\Re e \mathcal{E}$, and $\Re e \widetilde{\mathcal{H}}$. 
    \item The DVCS cross section has minimal kinematic dependence, is flat in the azimuthal angle $\phi$, and contains all eight CFFs.
    \item $\Gamma = \displaystyle\frac{\alpha^3}{16\pi^2 (s-M^2)^2 \sqrt{1+\gamma^2}\, x_{Bj} } $ \cite{Kriesten:2019jep}
\item The coefficients $A_{\mathcal{I}}$, $B_{\mathcal{I}}$, and $C_{\mathcal{I}}$ have been calculated in Ref.\cite{Kriesten:2019jep} and their numerical values in specific kinematic bins are plotted in Ref.~\cite{Kriesten:2020wcx}.
\item  The cross section is presented at leading order. Other terms of twist-three can in principle enter but they are expected to be suppressed in the asymptotic regime, $Q^2>>t, M^2$. We leave the subject of higher twists to a future study.
\end{itemize}

\subsection{Machine Learning Framework}
\label{sec:mlframework}

The use of neural network architectures in nuclear and particle physics
has being growing rapidly  (see, for example Refs.~\cite{Carleo:2019ptp,Rajput:2023swq,Bourilkov:2019yoi}).
The majority of these studies has been dedicated to experimental questions including, for instance, the optimization of detector performance and the kinematic reconstruction of scattering events. More recently, ML has been emerging as a possible framework for theoretical studies. The present analysis is born in this context: in this manuscript we show how ML methods play a critical role in the phenomenological analysis of multi-particle reactions such as deeply virtual exclusive scattering processes, allowing us to separate the physically meaningful information  from the unwanted features of the experimental data. The extracted information, expressed in terms of several observables, can then be quantified with its uncertainty. 

The robust nature of neural networks makes them optimal for this type of complex global analysis, allowing spaces for physics input into their architectures. 
%
Neural networks are first and foremost a statistically and mathematically rigorous tool for function approximation using layers of linear transformations passed through non-linear activation functions to learn the mathematical mapping from input data to predictions. The minimization of a loss function like the mean-squared-error, or the squared L2 norm, mimics the chi-square minimization techniques of industry standards such as Hessian-based analyses. 
In this paper we will focus on variational autoencoders (VAEs) built with multi-layered perceptrons (see Appendix~\ref{appendixB} for a general description).


\subsubsection{Variational Autoencoder}

A Variational Autoencoder (VAE) is a generative model extension of the Autoencoder (AE) that can be used for unsupervised learning of complex data distributions~\cite{vae, 2019arXiv190602691K}. It can learn a compact, low-dimensional representation of a dataset, and then use this representation to generate new data points that are similar to the original (see Appendix~\ref{appendixC} for a brief description of the AE).

\begin{figure}[t]
    \includegraphics[width=\linewidth,height=7.5cm]{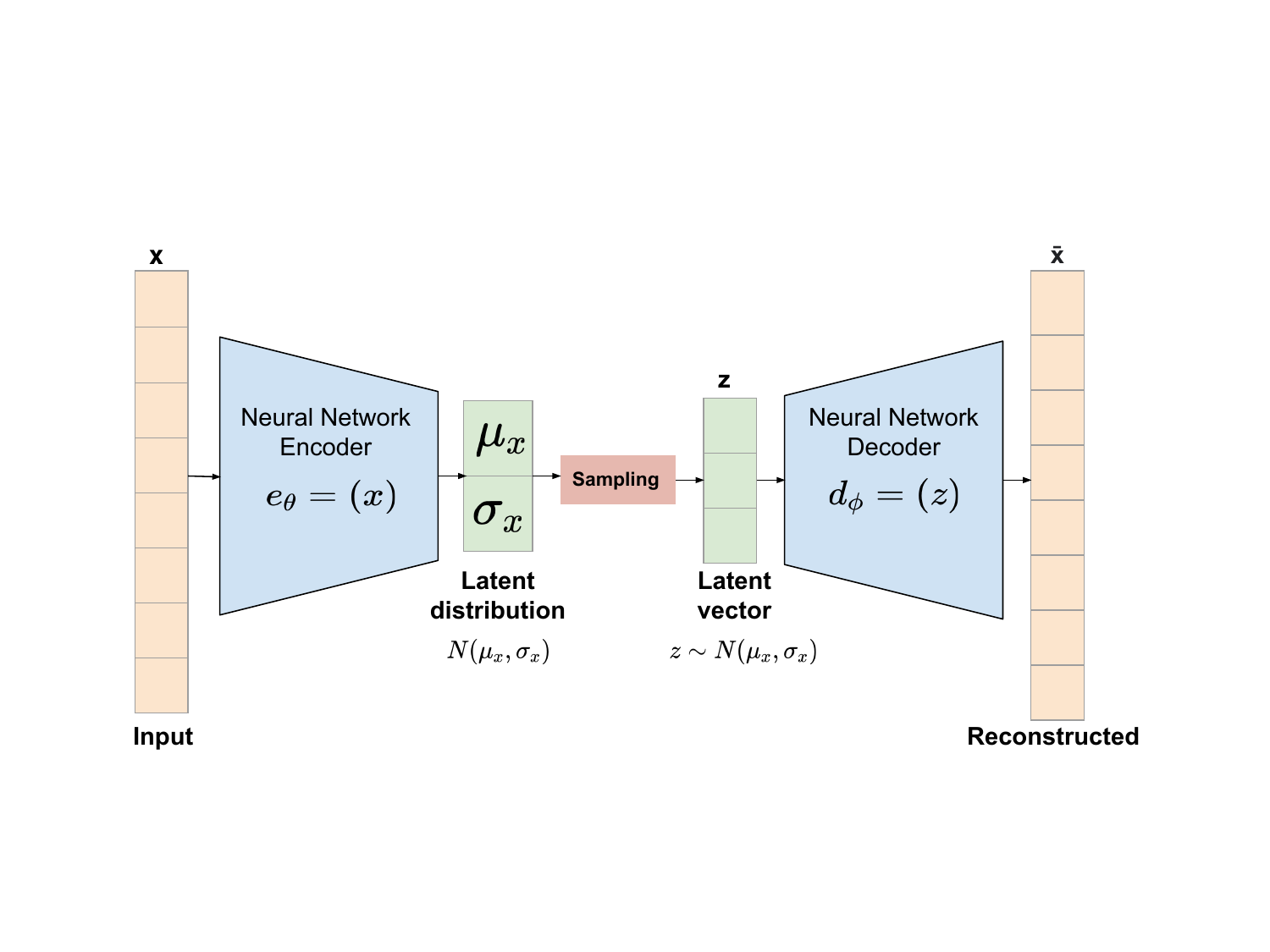}
    \vspace*{-20mm}
    \caption{General architecture of the variational autoencoder}
    \label{fig:vae_arch}
\end{figure}

As illustrated in Fig.~\ref{fig:vae_arch}, the VAE consists of two main components: an encoder and a decoder. The encoder of a VAE is an MLP which maps the input data $x$ to a latent representation $z$ through a function 
$e_{\theta}(x)$. $e_{\theta}(x)$ is, therefore given by a series of matrices. 
The latent representation $z$ is typically a vector of real numbers, and it is often constrained to follow a Gaussian distribution with mean ${\mu}$ and variance ${\sigma}$. The decoder of a VAE maps the latent representation $z$ back to the original space through a function 
$d_{\phi}(z)$. The VAE involves optimization of a loss function that consists of two terms: a reconstruction loss and a regularization term. \\

Loss = Reconstruction Loss + Regularization Term \\

The reconstruction loss measures the difference between the input data and its reconstruction, and it is usually calculated using the mean squared error(MSE).
The regularization term ensures that the latent representation, $z$, follows a specified distribution, such as a Gaussian distribution. As we explain in detail in Sec.~\ref{sec:vaim_background}, this term is calculated using the Kullback-Leibler (KL) divergence between the latent distribution and the specified distribution. 
%
This constraint allows the VAE to learn a continuous and structured latent space, which makes it better suited for tasks such as generating new data or performing interpolation between data points. 

The VAE is trained by minimizing the loss function using gradient descent or a similar optimization algorithm. This involves adjusting the parameters of both the encoder and the decoder to learn a good representation of the data.
Once the VAE is trained, it can be used to generate new data points by sampling from the latent distribution and passing the samples through the decoder. The generated data points will be similar to the original data, since they are constructed using the same underlying structure learned by the VAE.

\subsection{Uncertainty Quantification}
\label{sec:uq}

Deep neural networks and other machine learning architectures have quickly become an increasingly applicable tool in physics analysis; therefore, it is important to quantify how much we can trust ML predictions. In essence, the physics that you are trying to extract from a deep learning model has to lie within the error bands of the network's predictions.

There are two main sources of uncertainty that are propagated in a neural network analysis. Epistemic uncertainty, a type of reducible error, arises from choices made in the framing of the problem and the construction of the neural network architecture. Examples include choice of depth of the model, activation functions used, number of parameters, model simplifications, limited training data. There is also aleatoric uncertainty, an irreducible error, which is encapsulated as data uncertainty. Examples of aleatoric uncertainty include statistical uncertainties in reported measurements, systematic uncertainties from the experimental techniques themselves, as well as generalization uncertainties to out of domain regions from the training data.  


In this manuscript, we use the Monte Carlo dropout technique for UQ. Given a training dataset $\mathcal{D}$ that consists of some feature space ${\bf {x}}_{i}$ and some labels $y_{i}$, we can write the probability of the outcome of our neural network prediction as:
\begin{eqnarray}
    p(y | {\bf{x}}_{j},\mathcal{D}) = \int d w_{\theta} \,\,p(y | {\bf{x}}_{j}, w_{\theta}) \,p(w_{\theta}|\mathcal{D}).
\end{eqnarray}
The weights and biases of the neural network $w_{\theta}$ are tuned during training in order to construct the posterior $p(w_{\theta}|\mathcal{D})$. In Ref.~\cite{gal2015dropout} a connection was made between using Monte Carlo dropout during prediction and the construction of the posterior distribution in Bayesian neural networks. By using Monte Carlo sampling of the model's weights, the total uncertainty is approximated by the variance in the predictions. Other methods for UQ include Gaussian Regression Proceesses and Bayesian Neural Networks. 











\section{Variational Autoencoder Inverse Mappers for CFFs (VAIM-CFF)}
\label{sec:vaim_background}

Extracting the Compton form factors from the cross sections is an inverse problem which seeks to estimate the unknown parameters from measured observables. 
Most often the inverse problems are ill-posed, where the solution is non-unique and small distortions in the data can significantly influence the characteristics of the solutions. 
Hence, solving inverse problems through analytical approaches is challenging. 

With the recent advances of machine learning, deep learning approaches have become state-of-the-art in solving various inverse problems. 
Applied to several inverse problems in the science and engineering fields, deep learning based methods have shown robust performance compared to traditional numerical solution methods~\cite{Arridge2019SolvingIP}. 
For example, invertible neural networks (INNs) is proposed~\cite{ardizzone2018analyzing} to solve inverse problems by learning the forward mapping. As well, the mixture density network (MDN)~\cite{Bishop94mixturedensity} is utilized to solve inverse problems by constructing the conditional probability distribution of the parameters, given the observables.
To extract the Compton form factors (CFF), we adopt the Variational Autoencoder Inverse Mappers (VAIM) framework, focusing on approximating the posterior distributions of the CFF with respect to the given unpolarized cross section and kinematics. 
Past work in Ref.~\cite{Almaeen2021} demonstrates the success of applying VAIM in several inverse problems with different solutions pattern.




\subsection{ VAIM Architecture}
\label{sec:vaim_arch}

\begin{figure}[http]
    \includegraphics[width=\linewidth, height=7cm]{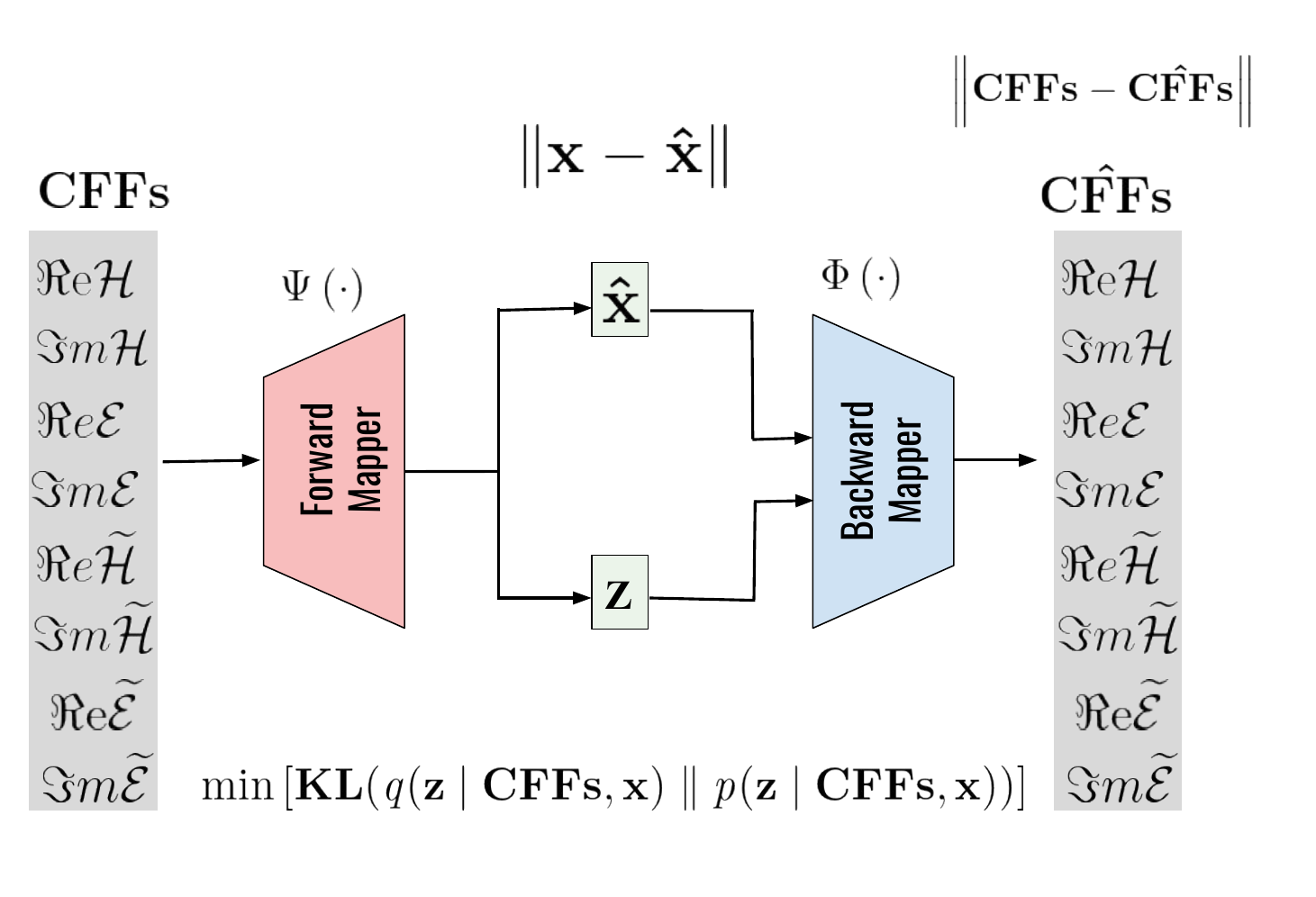}
    \caption{Overview of VAIM architecture}
    \label{fig:vaim_arch}
\end{figure}

VAIM architecture is illustrated in Fig.~\ref{fig:vaim_arch} consisting of two networks: a forward mapper $\Psi(\cdot)$ and a backward mapper $\Phi(\cdot)$. The forward mapper $\Psi(\cdot)$ maps the CFFs to the cross sections and latent variable $z$ and the backward mapper $\Phi(\cdot)$ maps the coupled cross sections and the latent variable $z$ to the CFFs. 

A latent variable is a variable that forms a part of the model, which it can not be directly observed, and thus is not part of our dataset. 
Typically, $z$ is used to denote such latent variables. 
Because of the information loss problem associated with the inverse problems, the latent variable $z$ is incorporated between the forward and backward mappers to capture the lost information and learn the patterns of the CFFs distributions. 
For efficient sampling, $z$  is restricted to follow well-known distributions, such as Gaussian or uniform distributions. Once VAIM is trained, sampling $z$ in the backward mappers leads to the corresponding CFFs distribution.

Indicating the Compton form factors as $\mathbf{cff}\in\mathbb{R}^m$, the cross sections as $\mathbf{x}\in\mathbb{R}^n$, and the latent variables in the latent layer as $\mathbf{z}\in\mathbb{R}^k$, the forward mapper learns the one-to-many mapping from CFFs to the cross sections and the latent $z$ such that 
    $$\Psi(\mathbf{cff}; w_f) = [\mathbf{\hat{x}}, \mathbf{z}],$$
\noindent where $w_f$ represents the weights of the forward mapper neural network, and $\mathbf{\hat{x}}$ is an approximation to the observed $\mathbf{x}$. 
The backward mapper is trained to reconstruct $\mathbf{cff}$ according to $\mathbf{x}$ and $\mathbf{z}$ such that
    $$\Phi(\mathbf{x}, \mathbf{z}; w_g) = \mathbf{\hat{cff}},$$ 
where $w_g$ are the weights of the backward mapper neural network, and $\mathbf{\hat{cff}}$ approximates the Compton form factors $\mathbf{cff}$.

 \subsection{VAIM Model}
 \label{sec:vaim_model}
 
In VAIM, an inverse problem is represented as a statistical inverse problem. With the given observed cross sections, VAIM learns to approximate a probability distribution of the CFFs. The forward mapper learns to approximate the posterior distribution 
    $p(\mathbf{z}\,|\,\mathbf{cff}, \mathbf{x})$ by another distribution
    $q(\mathbf{z}\,|\,\mathbf{cff}, \mathbf{x})$,
and the backward mapper learns to approximate the likelihood distribution $p(\mathbf{cff}, \mathbf{x} \mid \mathbf{z})$. 

Since $p(\mathbf{z}\,|\,\mathbf{cff}, \mathbf{x})$ is intractable, the variational inference \cite{variationalinference} is used to approximate the true posterior distribution $p(\mathbf{z}\,|\,\mathbf{cff}, \mathbf{x})$
by learning another tractable distribution 
    $q(\mathbf{z}\,|\,\mathbf{cff}, \mathbf{x})$. 
VAIM model is optimized to ensure that these distribution are nearly equivalent
    $$q(\mathbf{z}\,|\,\mathbf{cff}, \mathbf{x})
    \sim p(\mathbf{z}\,|\,\mathbf{cff}, \mathbf{x}),$$
A typical metric that is often used to quantify the difference between two probability distributions is The Kullback-Leibler (KL) divergence. Where the minimization of the (KL) divergence, leads to the two distributions $q$ and $p$ are being similar. VAIM is then modeled as 
    $$\min \Big[ {\rm KL}\big( q(\mathbf{z}\,|\,\mathbf{cff}, \mathbf{x})
    \mid\mid p(\mathbf{z}\,|\,\mathbf{cff}, \mathbf{x}) \big) \Big],$$
where
    ${\rm KL}\big( q(\mathbf{z}\,|\,\mathbf{cff}, \mathbf{x})
    \mid\mid p(\mathbf{z}\,|\,\mathbf{cff}, \mathbf{x}) \big)$
denotes the KL-divergence between $q$ and $p$. With the variational autoencoder (VAE) theory given in \cite{vae}, the previous optimization problem is equivalent to
    $$\min_{w_f, w_g} \Big[ \|\mathbf{x}-\mathbf{\hat{x}}\|_2^2 
    + \|\mathbf{cff}-\mathbf{\hat{cff}}\|_2^2 
    + {\rm KL}\big( q(\mathbf{z} \mid \mathbf{cff}, \mathbf{x}) \mid\mid p(\mathbf{z}) \big) \Big],$$
where $\| \cdot \|_{2}^{2}$ is the L$_{2}$ norm squared, $\|\mathbf{x}-\mathbf{\hat{x}}\|_2^2$ is the forward mapping error, $\|\mathbf{cff}-\mathbf{\hat{cff}}\|_2^2$ is the likelihood error of reconstructing the Compton form factors, and $p(\mathbf{z})$ is the true prior distribution. Often, $p(\mathbf{z})$ is selected to be a tractable, easy-to-generate distribution, such as a normal distribution or a uniform distribution. 
The KL-divergence approximation ensures that 
    $$q(\mathbf{z}\,|\,\mathbf{cff}, \mathbf{x}) \sim p(\mathbf{z}).$$

In the VAIM model, the latent variables $\mathbf{z}$ in the latent layer are the main component. The prior distribution $p(\mathbf{z})$ is tractable to approximate the true posterior distribution $p(\mathbf{z}\,|\,\mathbf{x}, \mathbf{y})$. 
 Throughout VAIM training, the latent layer seeks to learn the lost information in forward mapping and converts the ill-posed inverse problem into a well-posed regression problem in the backward mapper using the carried information in the latent space. 
 Specifying $z$ dimensions as greater or equal to the actual dimensionality of the information lost in forward mapping,
 the latent layer can reconstruct the complete Compton form factors distribution for the given observables.

\subsection{Conditional VAIM (C-VAIM)}

To expand CFFs extraction and include several sets of kinematics, we extend VAIM to the Conditional VAIM (C-VAIM) that is based on the standard C-VAE~\cite{NIPS2015_8d55a249}. With the C-VAIM we extend the number of kinematic variables that the VAIM can be trained on to include: $x_{Bj}$, $t$, $Q^{2}$. The values used in this paper are shown in Table~\ref{Table:dvcs kinematics values}.  

\subsubsection{C-VAIM architecture}

\begin{figure}[ht]
\includegraphics[width=\linewidth]{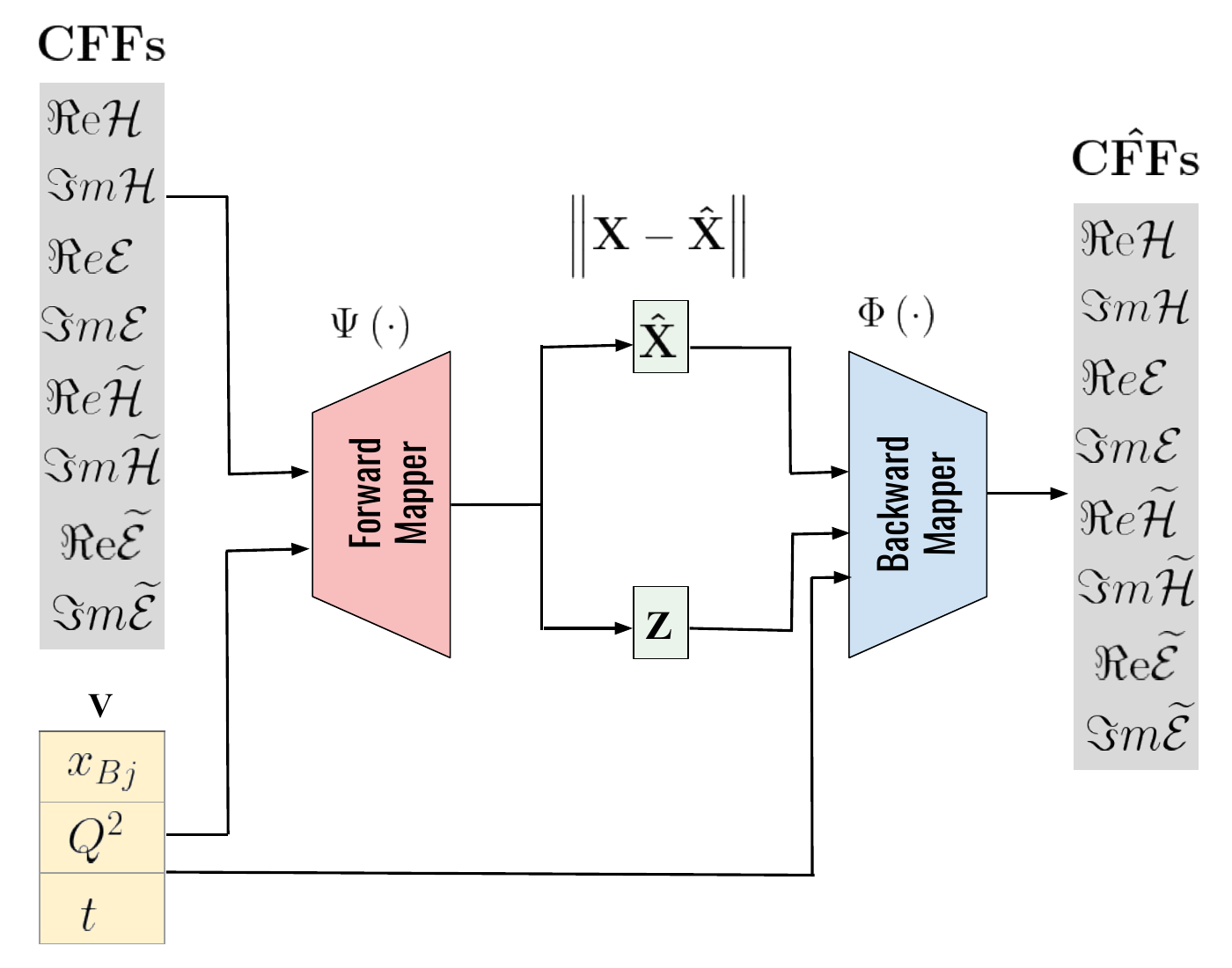}
\caption{Overview of C-VAIM architecture}
\label{fig:cvaim_arch}
\end{figure}

The overall architecture of C-VAIM is shown in Fig.~\ref{fig:cvaim_arch}, consisting of a forward mapper $\Psi(\cdot)$ and a backward mapper $\Phi(\cdot)$. The forward mapper $\Psi(\cdot)$ maps the CFFs and the kinematics to the cross sections and latent variable $z$ and the backward mapper $\Phi(\cdot)$ maps the coupled cross sections, the kinematics and the latent variable $z$ to the CFFs. 

C-VAIM  provides the advantage of imposing conditions over the forward and backward mappers. Thus, we can control the structure of the generated data during the backward mapping. In other words, with the training process on the entire kinematics data, we want our backward mappers to generate CFFs that correspond to a certain kinematics region. For this task, our VAIM assumption is slightly modified by conditioning the forward and the backward mapper to the kinematics $(x_{bj}, Q^2, t)$ denoted as $v$. In C-VAIM, the forward mapper attempts to approximate 
$p(\mathbf{z}\,|\,\mathbf{cff}, \mathbf{v}, \mathbf{x})$ by another distribution
    $q(\mathbf{z}\,|\,\mathbf{cff} ,\mathbf{v},\mathbf{x})$,
and the backward mapper learns to approximate the likelihood distribution $p(\mathbf{cff},\mathbf{v}, \mathbf{x} \mid \mathbf{z})$. 

Similar to VAIM, KL divergence is utilized in C-VAIM to minimize the divergence of the two distribution $q$ and $p$. Thus, C-VAIM is modeled as follows  
    $$\min \Big[ {\rm KL}\big( q(\mathbf{z}\,|\,\mathbf{cff},\mathbf{v}, \mathbf{x})
    \mid\mid p(\mathbf{z}\,|\,\mathbf{cff},\mathbf{v}, \mathbf{x}) \big) \Big],$$
The above optimization leads to the following revised objective function
    $$\min_{w_f, w_g} \Big[ \|\mathbf{x}-\mathbf{\hat{x}}\|_2^2 
    + \|\mathbf{cff}-\mathbf{\hat{cff}}\|_2^2 
    + {\rm KL}\big( q(\mathbf{z} \mid \mathbf{cff}, \mathbf{v}, \mathbf{x}) \mid\mid p(\mathbf{z}|\,\mathbf{v}) \big) \Big],$$
$\|\mathbf{x}-\mathbf{\hat{x}}\|_2^2$ represents the forward mapping error, $\|\mathbf{cff}-\mathbf{\hat{cff}}\|_2^2$ is the likelihood error of reconstructing the Compton form factors, and $p(\mathbf{z|\,\mathbf{v}})$ is the true prior distribution conditioned by the kinematics $v$. As VAIM, we choose $p(\mathbf{z})$ to be a tractable and easy-to-generate distribution, such as a normal distribution or a uniform distribution. 
As well the KL-divergence approximation to ensure that 
    $$q(\mathbf{z}\,|\,\mathbf{cff}, \mathbf{v}, \mathbf{x}) \sim p(\mathbf{z}|\,\mathbf{v}).$$ 

C-VAIM hyperparameters such as the number of neurons, number of layers, and the learning rate, are selected using Keras tuner~\cite{omalley2019kerastuner}, an open-source library to designate our architecture and perform hyperparameter tuning. The optimal hyperparameter is used as the final hyperparameter for our C-VAIM architecture. Our optimum forward mapper network architecture composed of an input layer whose nodes denote the array of the CFFs, followed by three fully-connected layers with 1024 neurons activated by a Leaky ReLU function. The output layer represents the cross section, kinematics and latent variables $z$. The network is regularized by an L2-norm penalty and a dropout rate of 0.2 to prevent overfitting. The backward mapper network architecture composed of an input layer that denotes the concatenated cross sections, kinematics with the latent layer $z$, followed by three fully-connected layers with 1024 neurons activated by a Leaky ReLU function. The output layer represents the reconstructed CFFs.

\subsection{CFFs Extraction}
\label{sec:cffs}

\begin{figure}[h]
\includegraphics[width=0.9\linewidth]{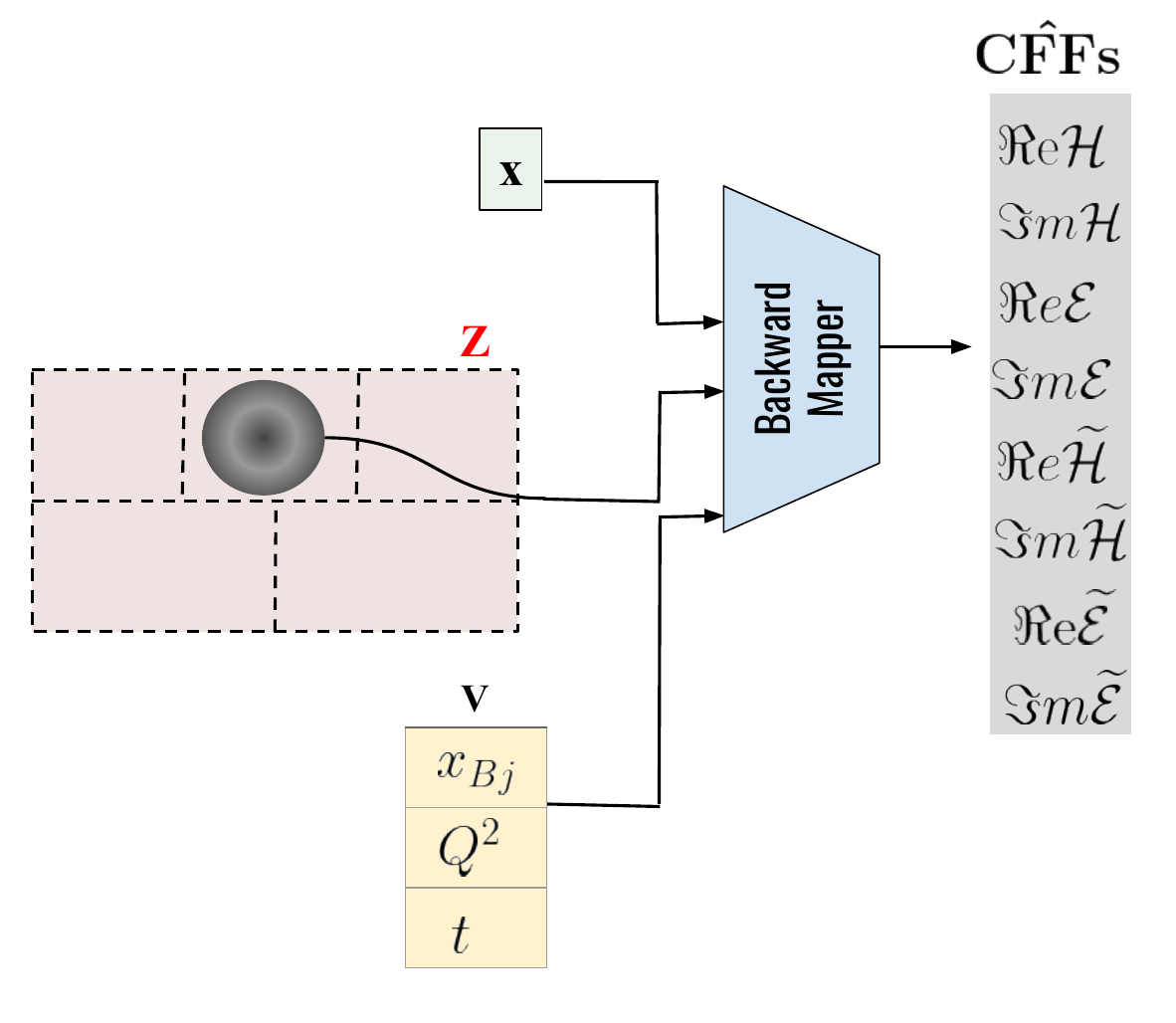}
\caption{VAIM backward mapper is used as a generative model to produce CFFs by sampling $\mathbf{z}$ with respect to the given cross section and kinematics.}
\label{fig:cff_backward}
\end{figure}

C-VAIM training process is a hybrid of supervised learning and unsupervised learning. 
The training of the forward mapper from CFFs and kinematics $v$ to the cross section, and the backward mappers from cross section to CFFs is supervised training. Whereas the derivation of the posterior distribution $p(\mathbf{z}\,|\,\mathbf{cff}, \mathbf{v}, \mathbf{x})$ is unsupervised training. 
The latent variable $z$ is distributed under the conditional probability distribution $p(\mathbf{z}\,|\,\mathbf{v})$. This means for every possible value of $v$, we will have a $p(\mathbf{z})$ that we can use to generate the CFFs corresponds to a specific kinematics value. 
After C-VAIM is trained, the backward mapper is used as a generative model to extract CFFs by predicting the combined cross sections and kinematics that are coupled with samples of the latent variable that is drawn from a specific $v$, as shown in Fig.~\ref{fig:cff_backward}.

\subsection{Procedure}
\label{sec:VAIMCFF_procedure}

We train C-VAIM on a set of kinematics values located in $x_{Bj} = [0.34,0.37]$, $t = [-0.17,-0.4]$ GeV$^{2}$, $Q^{2} = [1.82,1.99]$ GeV$^{2}$. The overall process of extracting CFFs is summarized as follow: 
\begin{enumerate}
    \item{Training region prior:} \\
 Generate a randomly distributed CFFs in  a specific ranges that are $\Re e \mathcal{H} = [-4,4]$, $\Re e \mathcal{E} = [-4,4]$, $\Re e \widetilde{\mathcal{H}}=[-10,10]$, $\Re e \widetilde{\mathcal{E}}=[-10,30]$, $\Im m \mathcal{H} = [-1,5]$, $\Im m \mathcal{E}=[-1,5]$, $\Im m \widetilde{\mathcal{H}}=[-1,20]$, $\Im m \widetilde{\mathcal{E}}=[-10,30]$. Compute the associated cross sections using physics calculations in Ref.~\cite{Kriesten:2019jep}.
    \item Training C-VAIM:\\
    With the generated CFFs and cross sections samples, we set up the forward mapper from the CFFs to the cross sections and the latent and the backward mapper from the cross sections and the latent to CFFs, as illustrated in Fig.~\ref{fig:cvaim_arch}.
     \item Prediction:\\
     Utilizing the C-VAIM backward mapper as a generative model, we test on the experimental cross sections that are concatenated with the latent space samples with respect to the kinematics.
     \item Validation:\\
     The predicted CFFs are validated by computing the corresponding cross sections using the theory calculation in \cite{Kriesten:2019jep}, and compare the cross section with the experimental cross sections.   
\end{enumerate}

\section{Results}
\label{sec:results}


In this section we show the results of our VAIM framework that is trained on the cross sections of several $x_{Bj}$, $t$, $Q^{2}$  kinematics values shown in Table~\ref{Table:dvcs kinematics values}. The C-VAIM is not limited to these values and can be easily extended to include all available kinematics. We leave this extension to a more complete set of kinematics to a comprehensive global analysis.

\begin{table}[http]
\begin{center}
    \begin{tabular}{ |c | c| c | c | }   
    \hline
     Bin & $x_{bj}$ & $t$ (GeV$^{2}$)& $Q^2$ (GeV$^{2}$)\\
    \hline 
    \hline
     1&  0.343& $-0.172$ & 1.820 \\
    \hline 
     2&  0.368& $-0.232$ & 1.933\\
    \hline
    3 & 0.375 & $-0.278$ & 1.964\\
    \hline
    4 &  0.379 & $-0.323$ & 1.986 \\
    \hline
    5 & 0.381 & $-0.371$ & 1.999\\ 
    \hline
    \end{tabular}
    \caption{Kinematics values that are used to train C-VAIM }
    \label{Table:dvcs kinematics values}
\end{center}
\end{table}

\begin{figure}[http]
\includegraphics[width=\linewidth]{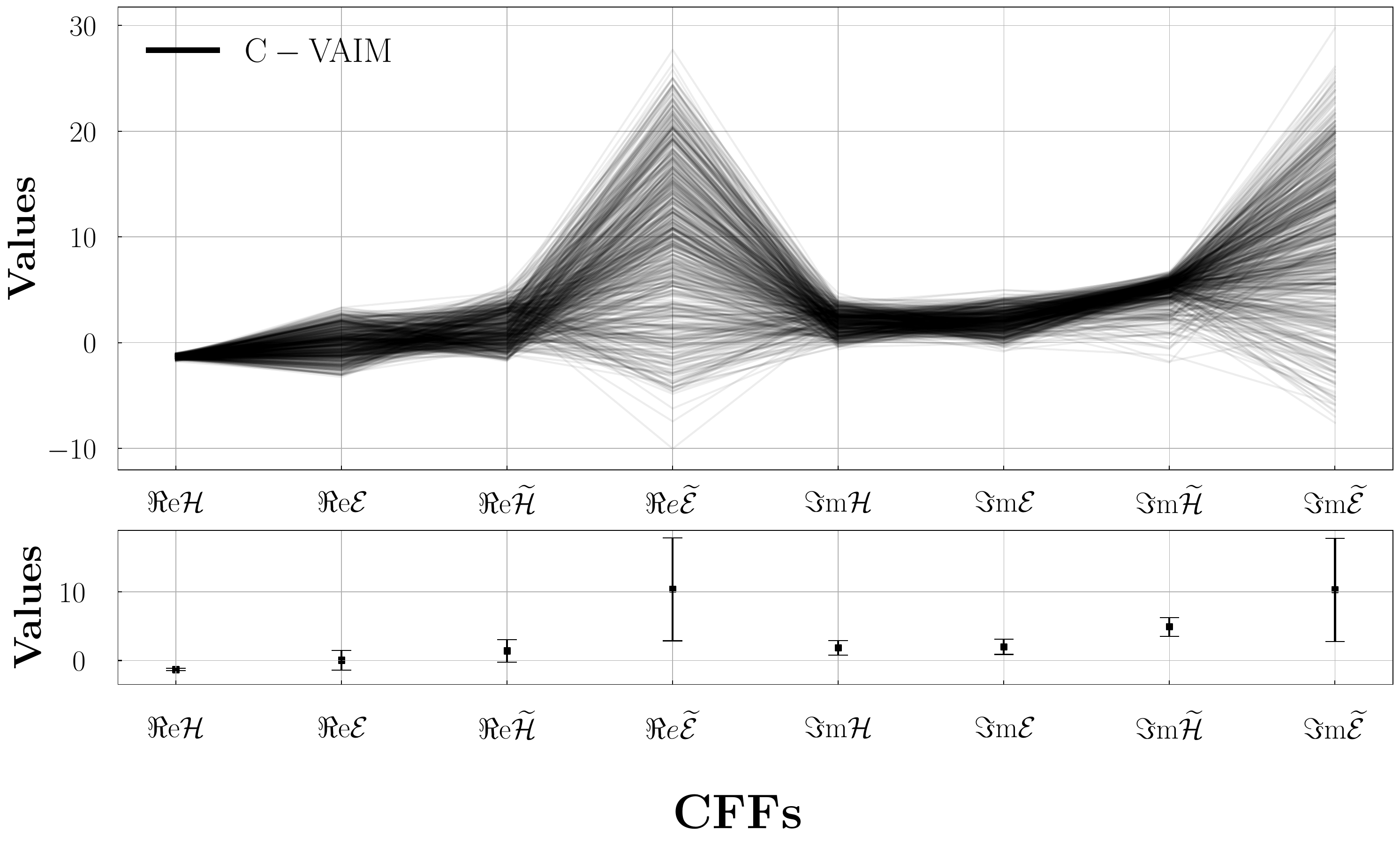}
\includegraphics[width=\linewidth]{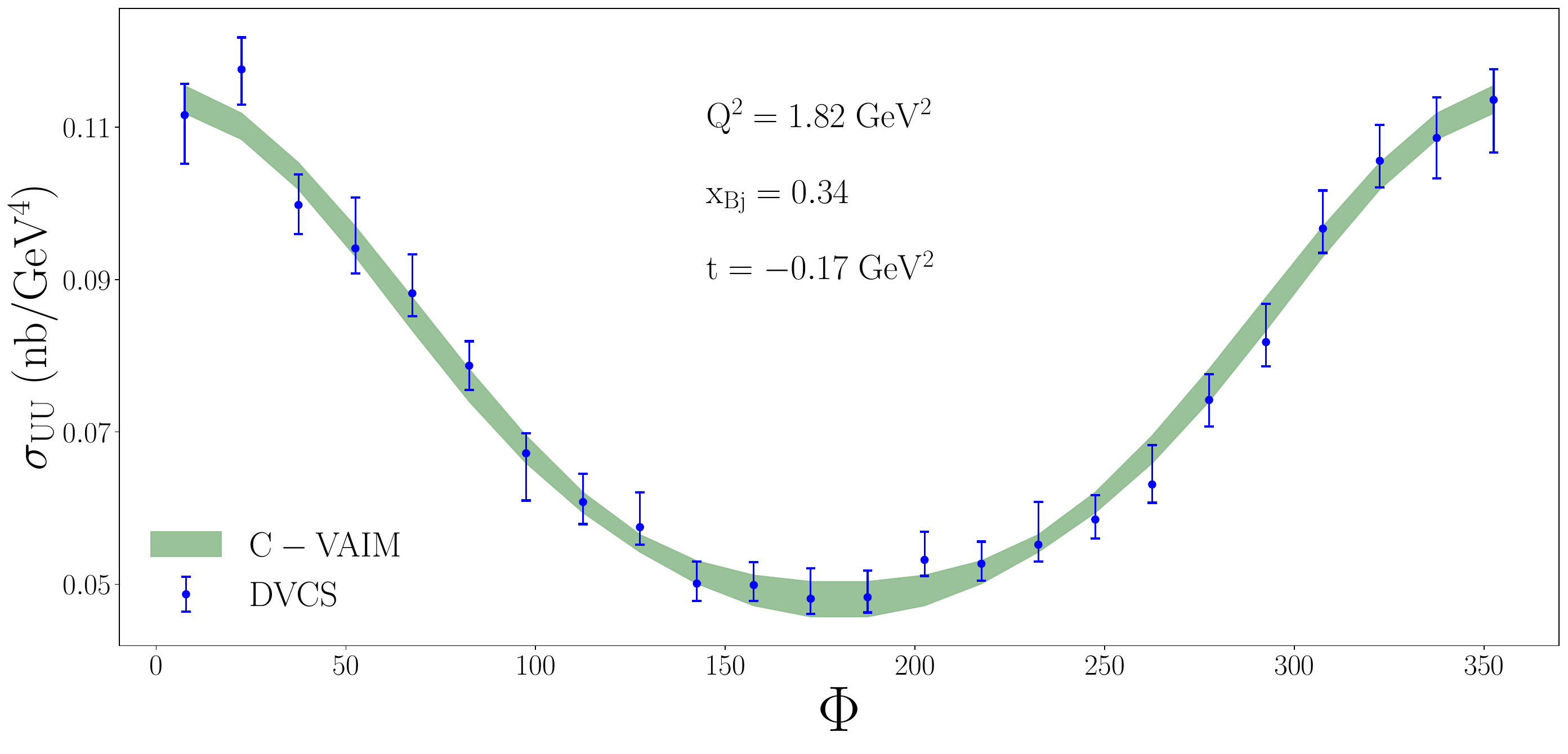}
\caption{(\textit{Top}) Predicted CFFs from the VAIM-CFF algorithm at a fixed kinematic experimental point $x_{Bj} = 0.343, t = 0.172$ GeV$^{2}$, and $Q^{2} = 1.820$ GeV$^{2}$. Predicted CFFs are generated in accordance with the training prior defined in Sec.~\ref{sec:VAIMCFF_procedure} and Ref.~\cite{JeffersonLabHallA:2015dwe}; 
(\textit{Middle}) The CFFs central values (blue squares) and standard deviation (bars around the blue squares) of the above solutions; 
(\textit{Bottom}) Ensembling the solutions as an uncertainty band, green, as compared to DVCS data, blue. }
\label{fig:PredCffs}
\end{figure}


In Fig.~\ref{fig:PredCffs} (top panel) we show a demonstration of the VAIM-CFF method in predicting the CFFs from a single kinematic point $x_{Bj} = 0.343$, $t=-0.172$ GeV$^{2}$, and $Q^{2} = 1.82$ GeV$^{2}$ of the unpolarized DVCS cross section at $5.75$ GeV beam energy; 
the  middle panel shows the central values and standard deviations of those predicted CFFs obtained by ensembling the results of the top panel; 
the bottom panel shows the predicted cross section plotted against experimentally measured data from Ref.~\cite{JeffersonLabHallA:2015dwe}.
We can point out a few characteristics of the analysis that one can infer from this extraction. The CFFs demonstrate a range of all possible solutions, with some CFFs being more well constrained than others by the widths of the prediction bands as compared to the widths of the training inputs. The CFFs are connected by lines to demonstrate correlations between predicted CFFs since all eight are predicted at the same time. 
The uncertainties of the experimental data are propagated into the CFF predictions through our method of random targets which we describe in a later section. Since there are no discernible ``multiple solutions" such as distinct groupings of banded solution sets, we take these solutions as an ensemble of Gaussian distributed predictions which we can then statistically treat to find a central value and a standard deviation as the error. These central values and errors are then propagated through the DVCS cross section and show as a band of cross sections as shown in Fig.~\ref{fig:PredCffs}(bottom) which fit the unpolarized cross section.

The extraction of $\Re e \mathcal{H}$ seems to be very well determined 
as it can be inferred from both the size of the ML model error and the statistical error from the experimental data shown in Fig.~\ref{fig:PredCffs}(middle). 
By ``well determined" we mean  that the predicted solution set from the VAIM-CFF is much narrower than the input uniform distribution. In the same vein, the unpolarized cross section does not seem to be able to narrowly constrain the CFFs $\Re e \widetilde{\mathcal{E}}$ nor $\Im m \widetilde{\mathcal{E}}$. We also see that the CFF from the longitudinally polarized GPD $\Re e$ and $\Im m$ $\widetilde{\mathcal{H}}$ are also rather well determined through the C-VAIM architecture applied to the unpolarized cross section. The fact that the unpolarized cross section only allows for a very specific range of $\Re e \mathcal{H}$ values, while not unexpected, is still surprising given the large spread found for the other CFFs.

\begin{figure*}[ht]
    \includegraphics[width=0.95\textwidth ]{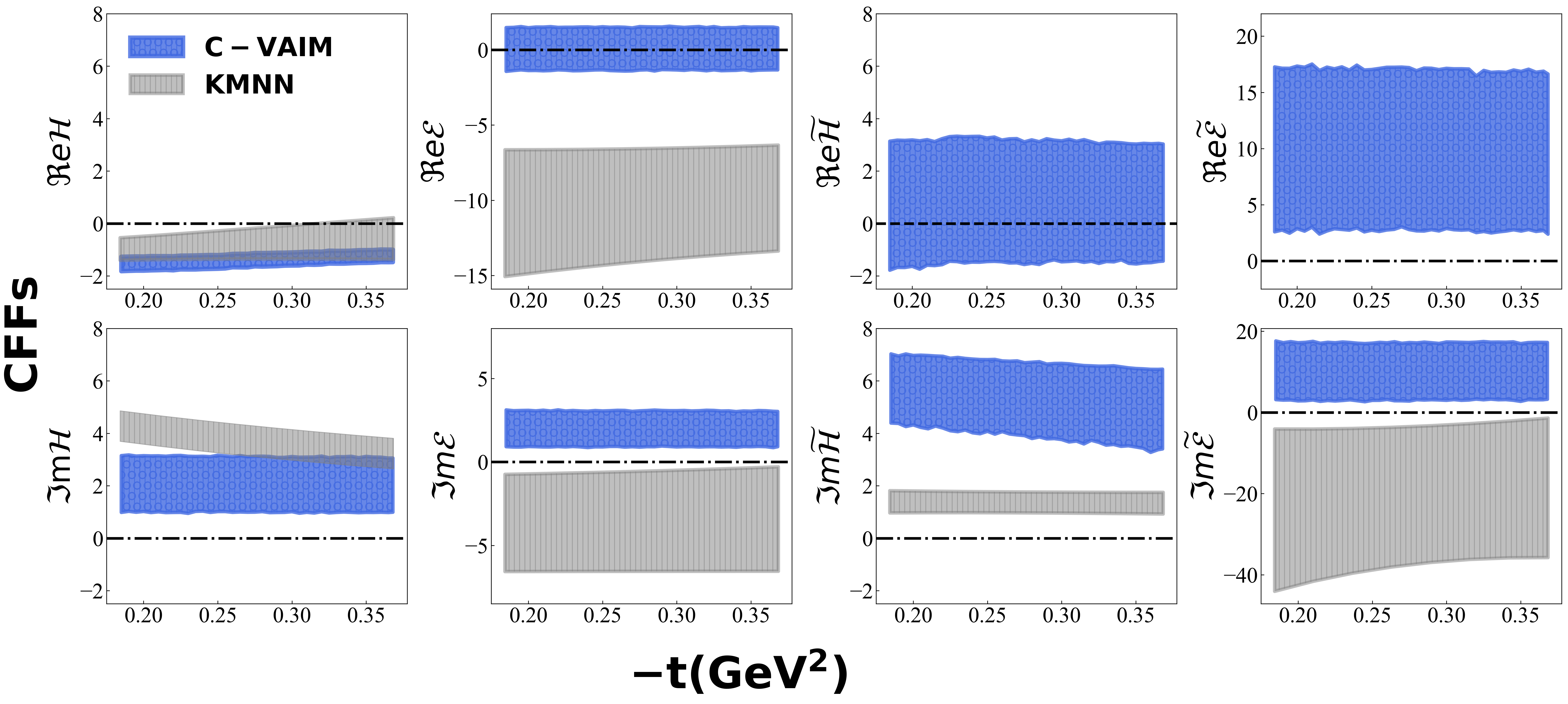}

    \caption{Prediction of all eight CFFs as a function of $t$ for a fixed kinematics: $x_{Bj} = 0.35$ and $Q^{2}=1.9$ GeV$^{2}$, and initial electron energy of 6 GeV~\cite{Defurne:2015kxq} 
    Results are compared with the NN based extraction of Ref.~\cite{Cuic:2020iwt}.
    Predictions are generated according to
    the training prior defined in Sec.~\ref{sec:VAIMCFF_procedure}.
    }
    \label{fig:cff_trends1}
\end{figure*}

\begin{figure*}[ht]
\includegraphics[width=0.95\textwidth]{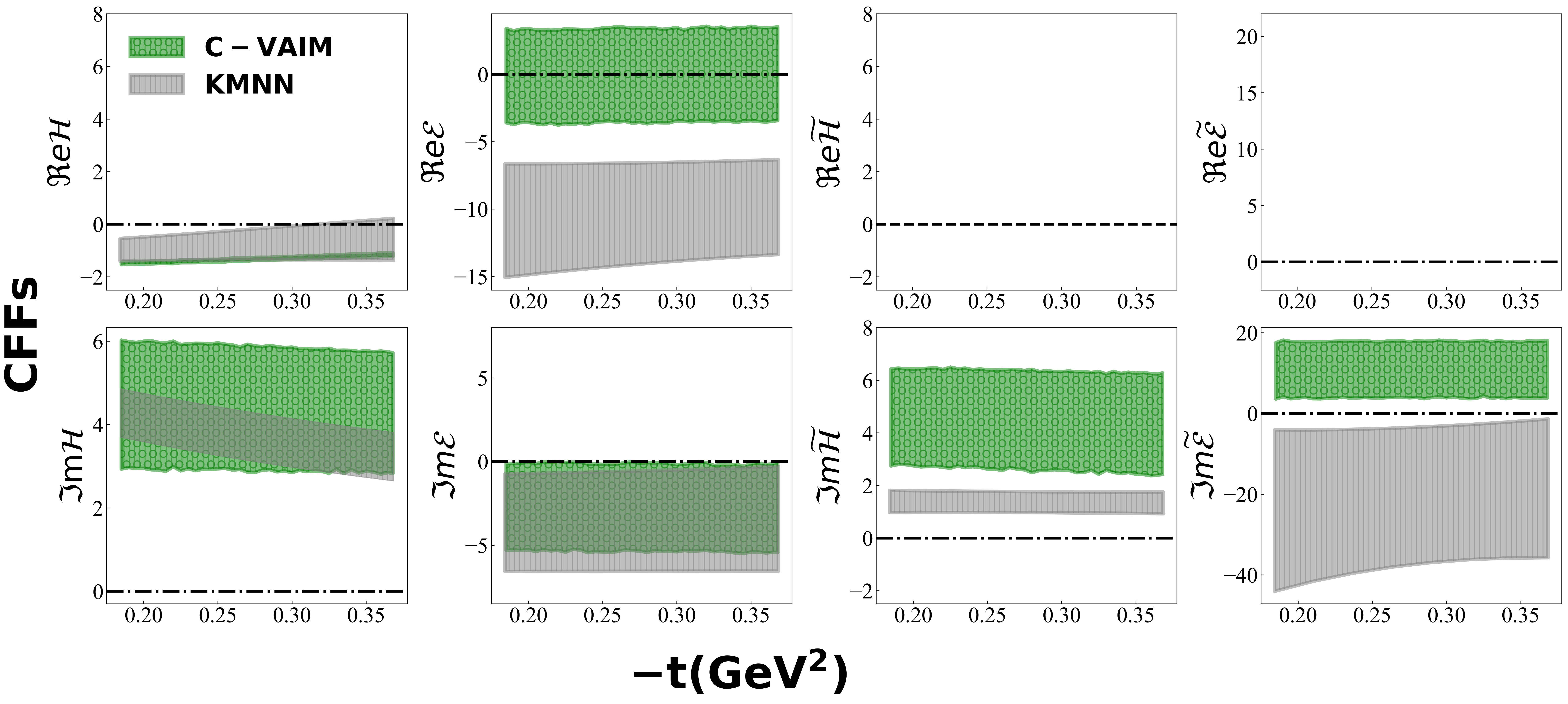}

      \caption{Prediction of six CFFs as a function of $t$ for a fixed kinematics $x_{Bj} = 0.35$ and $Q^{2}=1.9$ GeV$^{2}$ for 6 GeV initial electron energy \cite{Defurne:2015kxq} fixing $\Re e \widetilde{H}$ and $\Re e \widetilde{E}$ = 0. The choice of the six CFFs matches the one in the analysis of Ref.~\cite{Cuic:2020iwt}, therefore allowing for a more consistent comparison of results. 
    Predictions are generated according to the training prior defined in Sec.~\ref{sec:VAIMCFF_procedure}.
    }
    \label{fig:cff_trends}
\end{figure*}

In Fig.~\ref{fig:cff_trends1} we show predictions of the CFFs in a range of $t$ values for a fixed kinematics $x_{Bj} = 0.35$ and $Q^{2}=1.9$ GeV$^{2}$ for Jefferson Lab data at 6 GeV~\cite{JeffersonLabHallA:2015dwe}. We use a physics informed deep learning model developed in Ref.~\cite{Almaeen:2022imx} to interpolate the cross section in between experimental data points to feed into the C-VAIM architecture for predictions. Notice that the CFFs that were rather well determined in Fig.~\ref{fig:PredCffs} demonstrate noticeable trends in $t$. This is because the VAIM-CFF model can predict these CFFs with greater precision, the $t$-dependence of these three Compton form factors $ \Re e {\cal H}$ , $\Re e \widetilde{\cal H}$ and $\Im m \widetilde{\cal H}$ can be disentangled from the kinematic dependence. In Figure \ref{fig:cff_trends} we predict the $t$-dependence of the CFFs with only six of them entering into our fit (we exclude $\Re e \widetilde{H}$ and $\Re e \widetilde{E}$. The results of both figures are compared to another neural network extraction of CFFs~\cite{Cuic:2020iwt}. The six CFFs fit was performed to better compare to the extraction in  Ref.~\cite{Cuic:2020iwt} in which they predict the same CFFs using in a data range comparable to ours.



\begin{figure*}[ht]
    \includegraphics[width=\textwidth ]{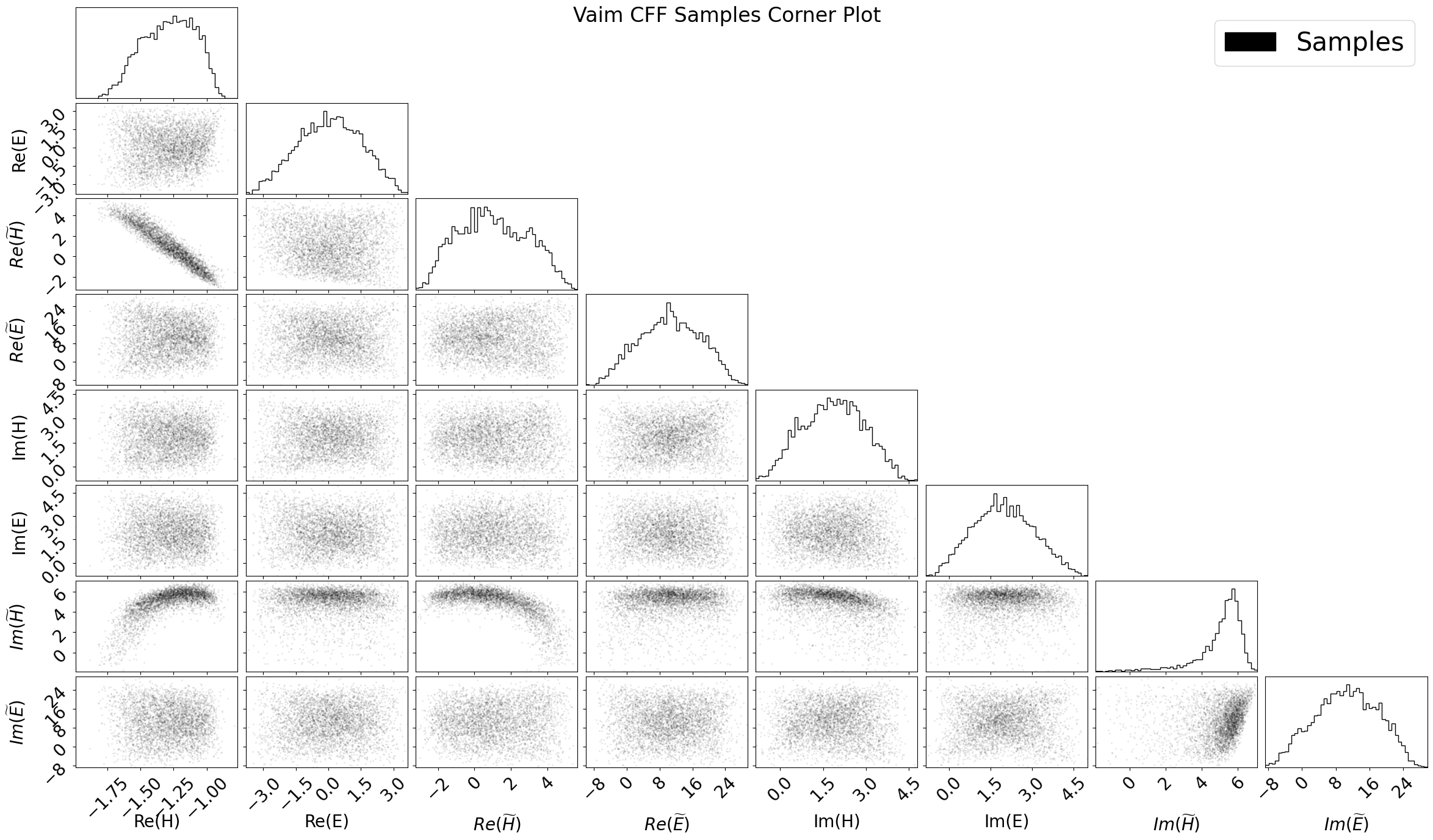}
    \caption{ Corner plot of VAIM CFF trials. Corner plot generated with corner.py tool~\cite{Mackey:2016joss}. CFF points are the same as those shown in other plots. The diagonal squares contain histograms. The off diagonal squares show scatter plot of samples generated by the VAIM algorithm. 
    Samples are generated with 
    $\xi=.207$,
    $x_{Bj}=0.343$,
    $Q^2  = 1.82$~GeV$^2$,
    $t=-0.172$~GeV$^2$.
    Samples are also in accordance with the training prior defined in Sec.~\ref{sec:VAIMCFF_procedure}.
    }
    \label{fig:cff_corner}
\end{figure*}

In Fig.~\ref{fig:cff_corner} we utilize a corner plot to study the correlations between the predicted CFFs. The use of the corner plot is to visualize, and emphasize the correlations from Fig.~\ref{fig:PredCffs} which are demonstrated by the lines connecting the predicted CFFs. 
Along the main diagonal are the histograms of the values of the CFFs, and the body are the predicted values of the CFFs plotted against one another. One can see that there are strong correlations between several of the CFFs (in particular $\Im m \mathcal{H}$ demonstrated by large structures of these points that display some sort of organization). 
Also notice the strong correlation between $\Re e \mathcal{H}$ and $\Re e \widetilde{\mathcal{H}}$. However; as is also noticed, there are significant numbers of points that are purely random - meaning that no correlation seems to exist between the predicted values. The information contained in this plot allows us to make key observations into the organization of the cross section.

We test the C-VAIM architecture on the cross sections as they exist in the literature (for a review of the differences see Ref.~\cite{Kriesten:2020wcx}). The cross sections as they are written are denoted as BKM~\cite{Belitsky:2010jw}, and UVA (labeled here as FemtoNet)~\cite{Kriesten:2019jep,Kriesten:2020wcx}. It is important to understand what effects the organization of the cross section has on the extracted observables (CFFs), as well as what information is contained in the cross section, and what information is generalized away in the measurements. We generate training samples using the two cross section formulations and train our C-VAIM independently. Trained C-VAIM decoder models are then applied on the DVCS cross section measurements for the kinematics $x_{Bj} = 0.343$, $t = -0.172$ GeV$^{2}$, and $Q^{2} = 1.820$ GeV$^{2}$. 

\begin{figure}[ht]
\centering
\includegraphics[width=\linewidth]{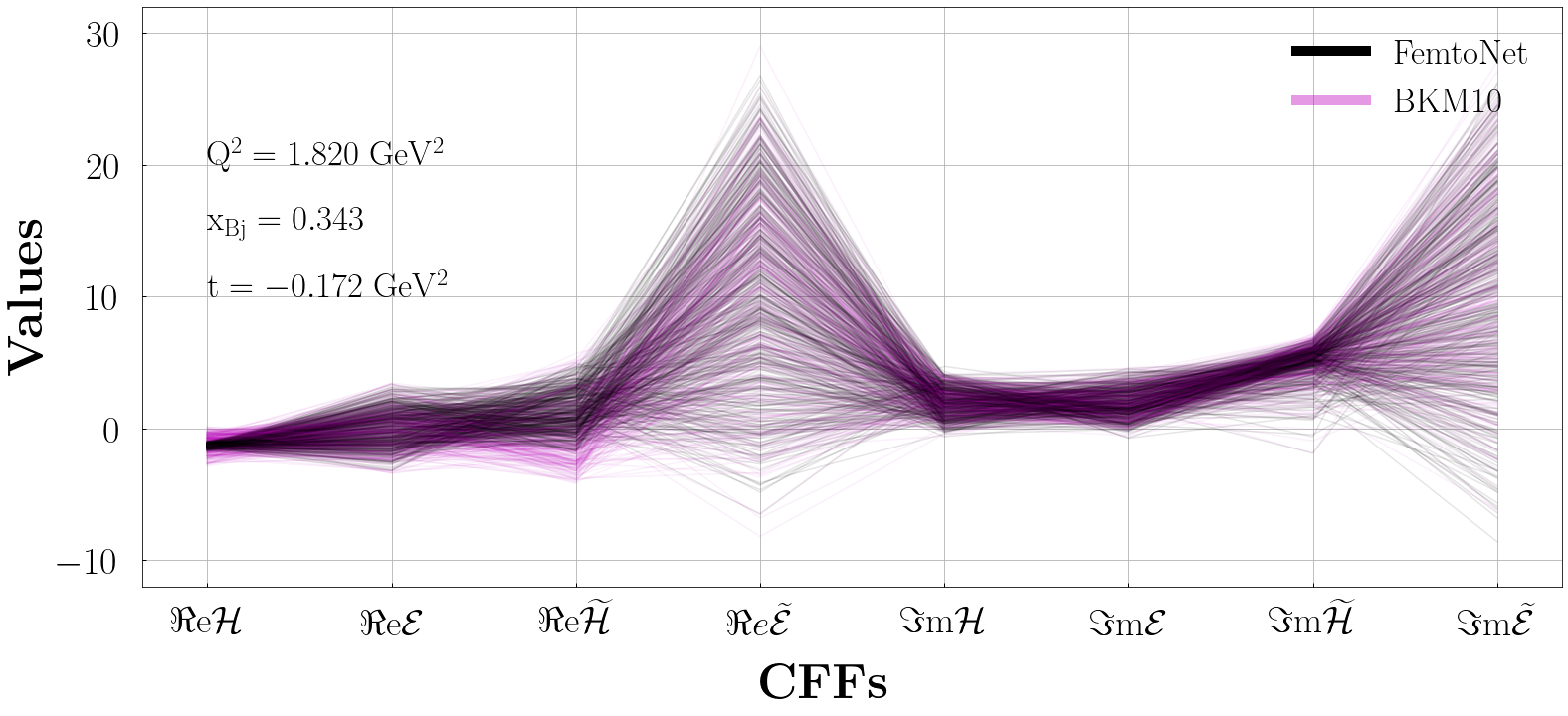}
\includegraphics[width=\linewidth]{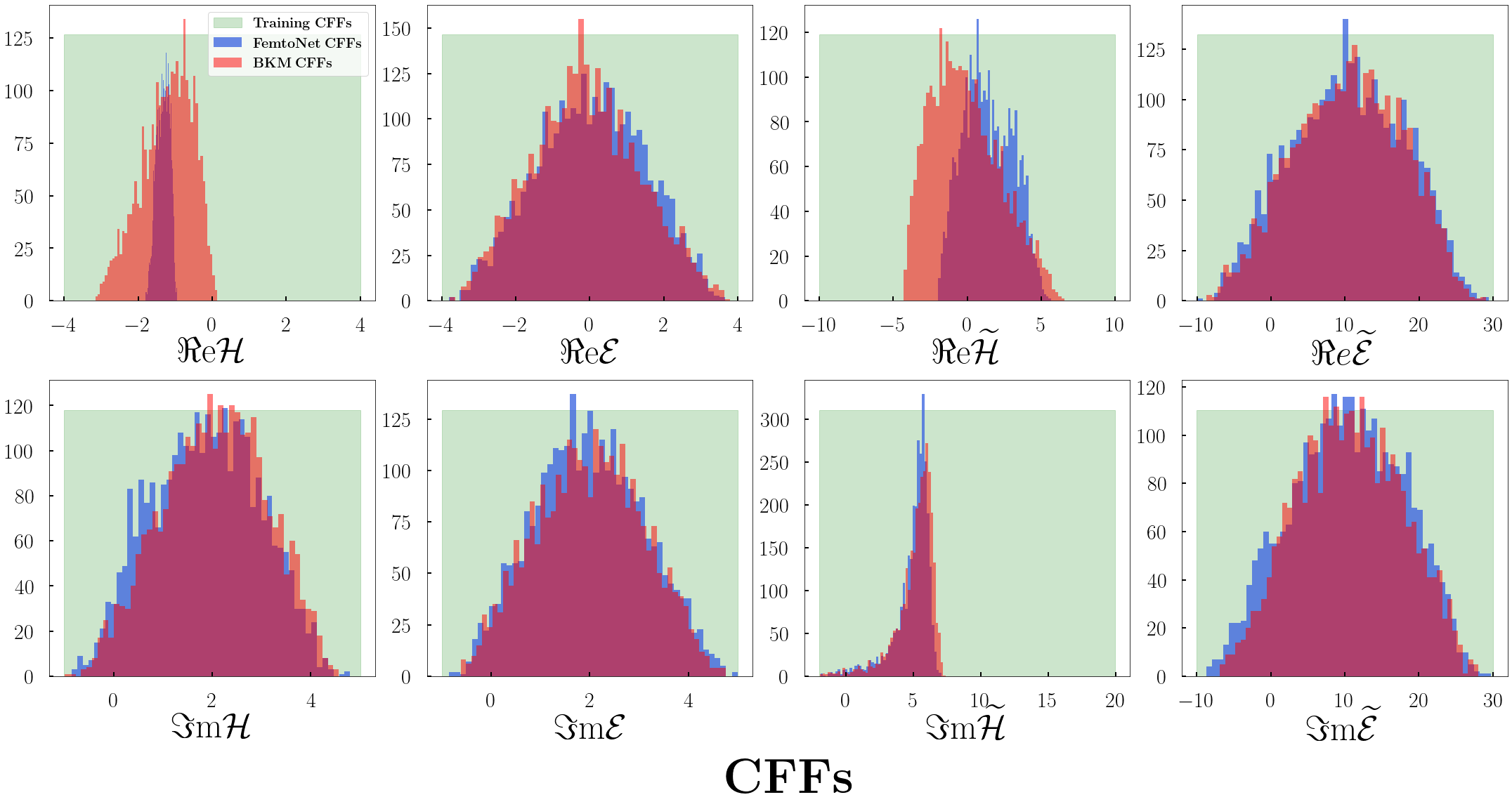}
\includegraphics[width=\linewidth]{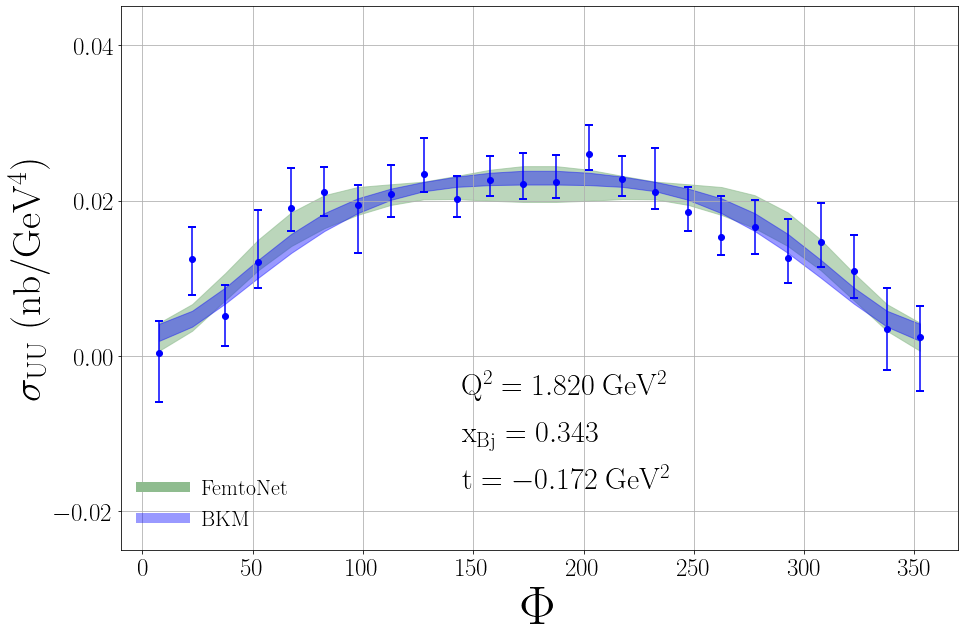}
\caption{Predictions of CFFs using C-VAIM framework comparing the cross section formulation of UVA \cite{Kriesten:2019jep,Kriesten:2020wcx} labeled FemtoNet and the formulation of Belitksy, Kirchner, and Mueller~\cite{Belitsky:2001ns,Belitsky:2010jw}. All results are shown for a specific kinematics for the unpolarized cross section at $x_{Bj} = 0.343, t = 0.172$ GeV$^{2}$, $Q^{2} = 1.820$ GeV$^{2}$, and $E_{b}=5.75$ GeV. (\textit{Top}) Comparison of outcomes of C-VAIM trained on the two different cross sections. 
(\textit{Middle}) Binning the outputs of C-VAIM as a histogram for each CFF to see the differences between BKM and UVA predicted CFFs. Also shown is the uniform input of CFF solutions for training. 
}
\label{fig:correct_NoCos}
\end{figure}

The predicted CFFs using the two formulations are overlapped as illustrated in Fig.~\ref{fig:correct_NoCos}, notice that there is not much difference between the predicted CFFs central values between BKM and FemtoNet; however, the uncertainty quantification seems to be much different. 
The histograms show this clearly where the CFFs predicted from BKM have a larger error as compared to the FemtoNet cross section. This indicates that the sensitivity to CFFs such as $\Re e \mathcal{H}$ is different between the two cross sections. 
We notice that the differences mainly appear in the interference terms 
 which are $\Re e \mathcal{H}$, $\Re e \widetilde{\mathcal{H}}$ where they have a noticeable appearance. Since the main organizational differences between the 2 cross section formulations is in the interference term, this is not surprising.
 
 We take the central values of the C-VAIM predicted CFFs and make predictions of the cross section from the fit. 
 We then compare both cross section formulations to the experimental data as reported with experimental uncertainties. 
 Notice that the larger uncertainties from the predictions leads to a smaller uncertainty band in the cross section, this is because particular CFFs like $\Re e \mathcal{H}$ seem to have less of a significant effect in one formulation than the other. 
 Since the CFFs only enter into the interference term and the DVCS squared contribution (and not the BH background) we subtract off the exactly calculable BH so that the differences are more easily noticeable. 
 It should be noted that in both instances we are fitting the CFFs to the data, so the total cross sections should overlap; however, the differences can mainly be seen in the extracted CFFs themselves.

\begin{figure}[h]
\includegraphics[width=\linewidth]{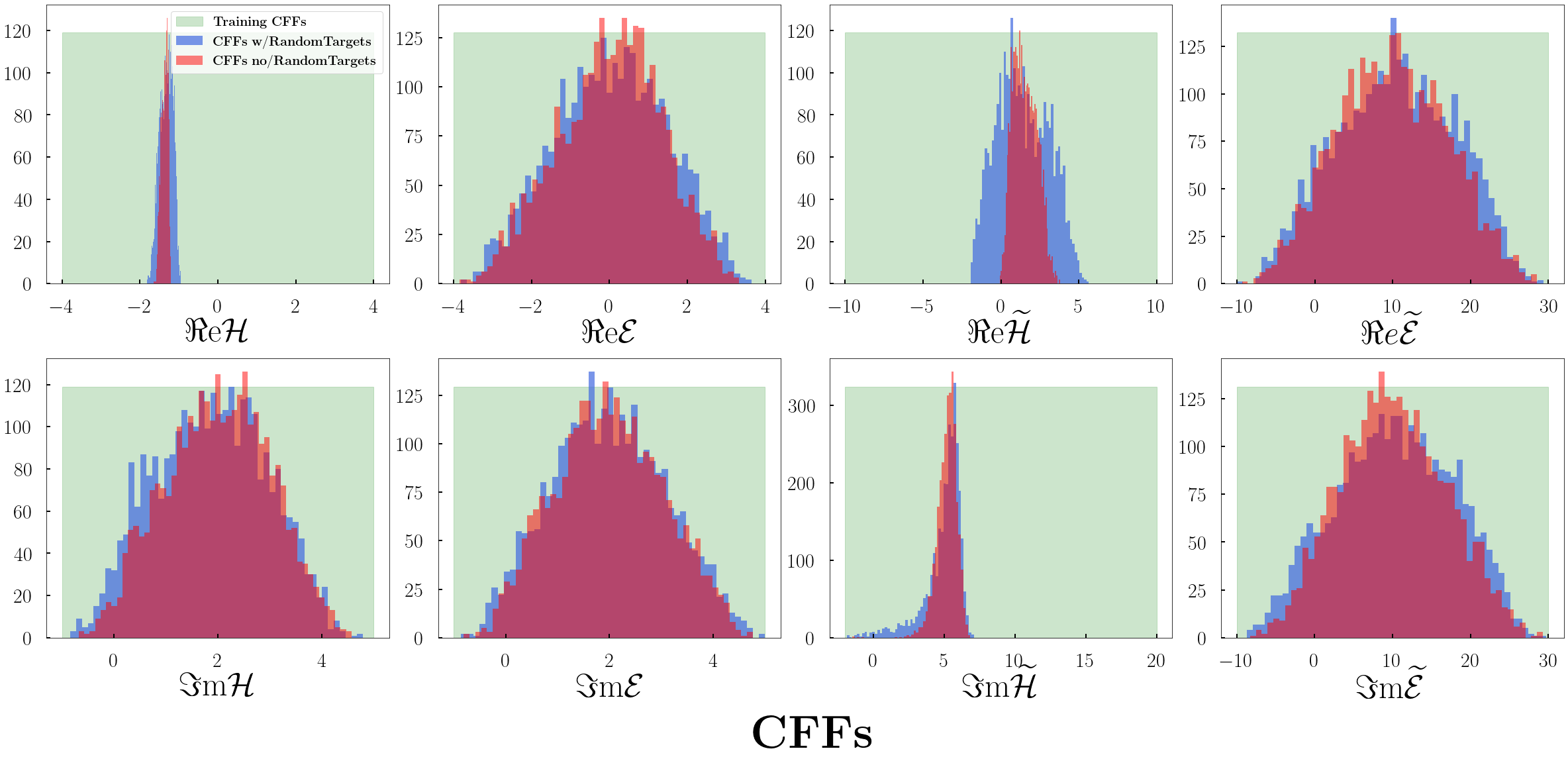}
\includegraphics[width=\linewidth]{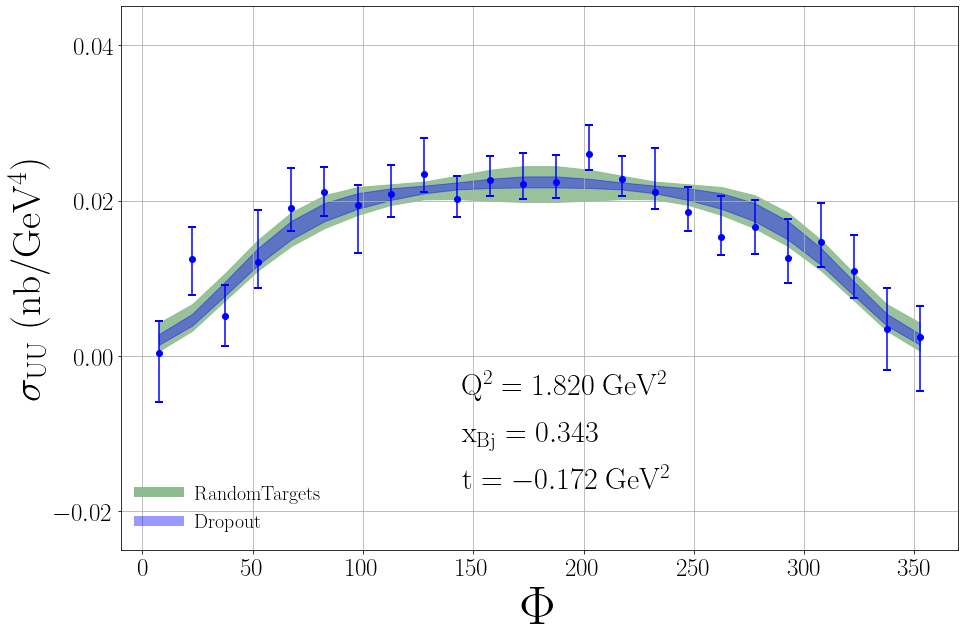}
\caption{Comparison of extracted CFFs at a particular kinematics $x_{Bj} = 0.343$, $t=-0.172$ GeV$^{2}$, $Q^{2} = 1.820$ GeV$^{2}$ using C-VAIM architecture both with and without the random targets propagated error.
(\textit{Top}) Comparison of the CFFs with and without the random  targets method to propagate uncertainty. 
(\textit{Bottom}) The uncertainty propagated through the DVCS cross section for both the interference term and DVCS squared term.}
\label{fig:correct_stErr}
\end{figure}

In Fig.~\ref{fig:correct_stErr} we compare the uncertainty associated to our predictions with and without the random targets method for the propagation of experimental uncertainties. 
This can mainly be seen in the presentation of the CFFs through histograms where we overlay the CFFs as predicted once with the random target and once with no random target. 
With no random target, the CFFs error bands are generated due to the variance of the solutions generated by C-VAIM through the latent variable (systematic errors from the C-VAIM architecture), and for the random targets method (combination of systematic from the C-VAIM and statistical errors from the experimental data). Notice that the random targets errors are consistently bigger, and making a more notable bandwidth on the three dominating CFFs which are $ \Re e {\cal H}$ , $\Re e \widetilde{\cal H}$ and $\Im m \widetilde{\cal H}$ suggesting that these CFFs are more sensitive to the statistical error (and more sensitive to the data in general). 
This also suggests that we are propagating the data uncertainties; however, that the error on the extracted CFFs are almost all contributions from reducible systematics.

\begin{figure}[ht]
\includegraphics[width=\columnwidth]{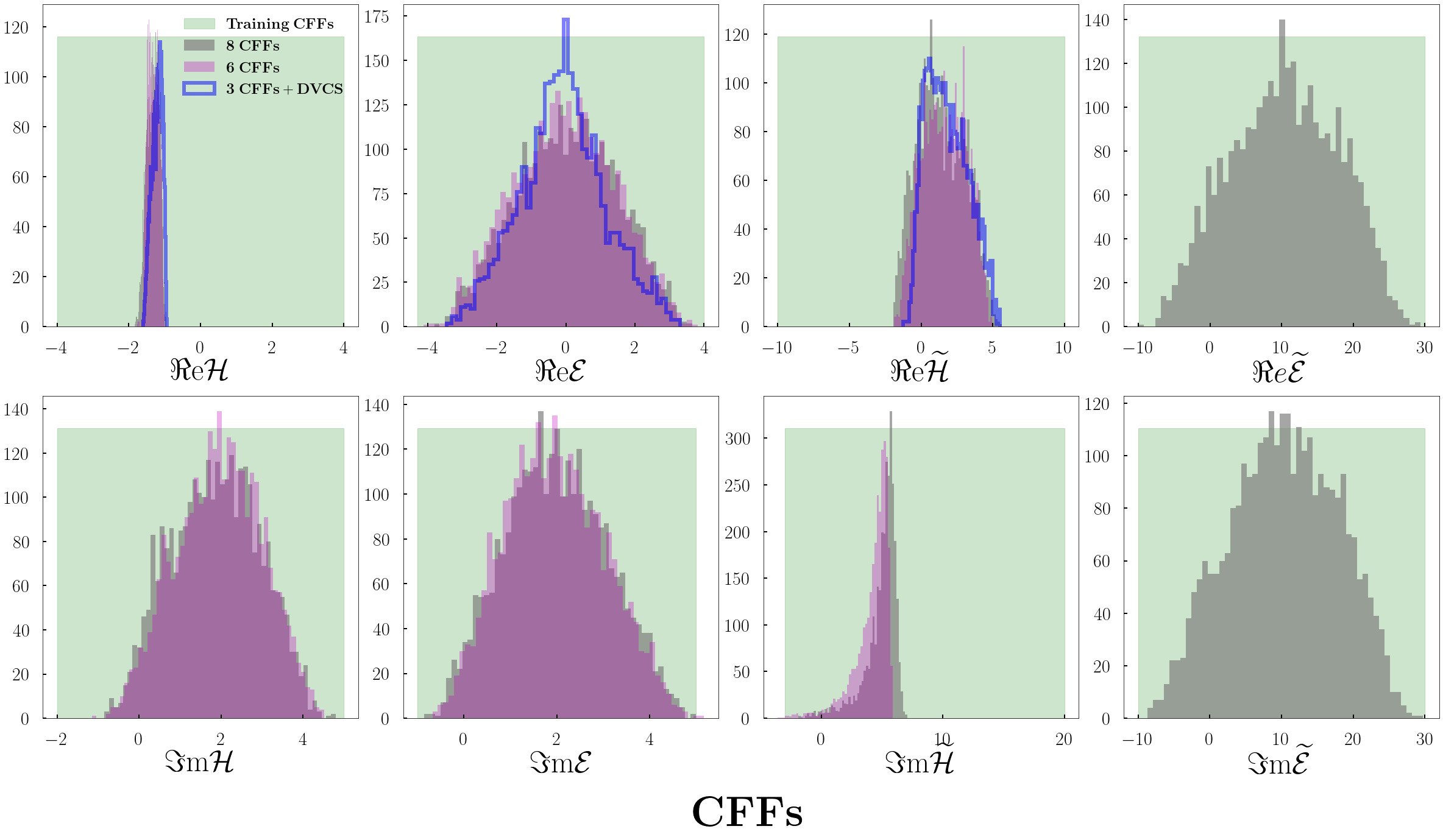}
\includegraphics[width=\columnwidth]{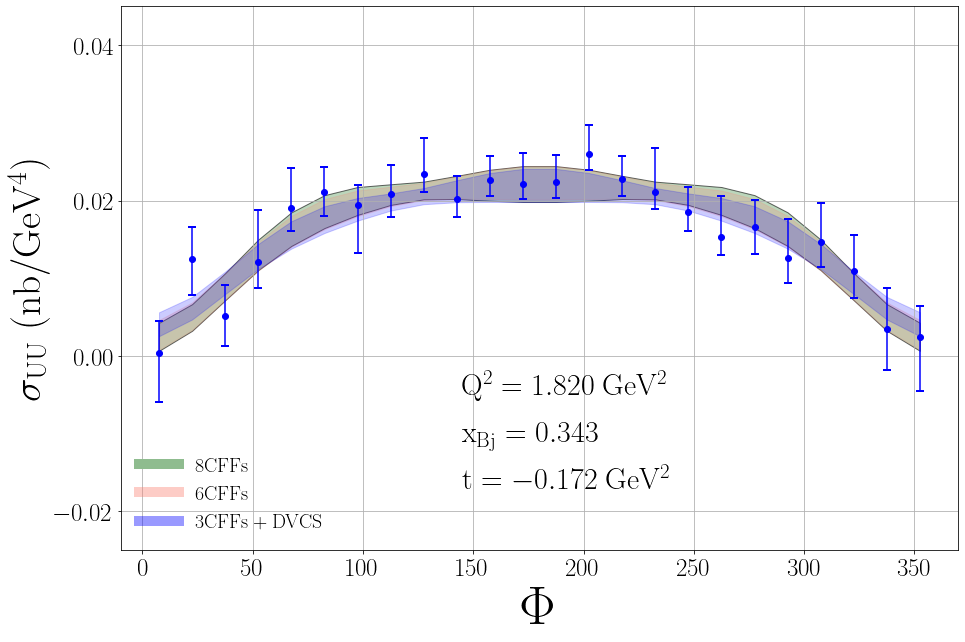}
\caption{Comparison of extracted CFFs based on organization of the cross section. Comparison of extracting all 8 CFFs, 6 CFFs, and 3 CFFs + DVCS cross section. At a kinematic value of $x_{Bj} = 0.343$, $t = -0.172$ GeV$^{2}$, $Q^{2} = 1.82$ GeV$^{2}$, and $E_{b} = 5.75$ GeV.}
\label{fig:3-6-8-CFFs-stErr}
\end{figure}

Using the structure of the DVCS cross section as formulated in Ref.~\cite{Kriesten:2019jep}, we study the sensitivity of this analysis to the number of extracted CFFs based on the cross section organization. Using the fact that the interference cross section is linearly dependent on three of the eight total CFFs, and is amplified with respect to the DVCS term by the interference with the BH background, we predict the three CFFs from the interference term and treat the total DVCS term as a parameter that C-VAIM can fit. 
This makes the extraction much simpler, in theory, as the C-VAIM no longer has to disentangle all eight CFFs from the DVCS term. One can find that the CFFs overlap in the three scenarios as demonstrated in Fig.~\ref{fig:3-6-8-CFFs-stErr}, with slight difference in the CFFs when using physics information from the structure of the DVCS cross section (3CFFs). 
We compare this result to an extraction of 6 CFFs from the unpolarized DVCS cross section, where we set $\Re e \widetilde{\mathcal{E}}$ and $\Im m \widetilde{\mathcal{E}}$ to zero. 
The results for 8 CFFs and 6 CFFs are nearly identical which indicates that the unpolarized DVCS cross section is not sensitive to $\Re e \widetilde{\mathcal{E}}$ and 
$\Im m\widetilde{\cal E}$.

\begin{figure}[ht]
\includegraphics[width=\linewidth]{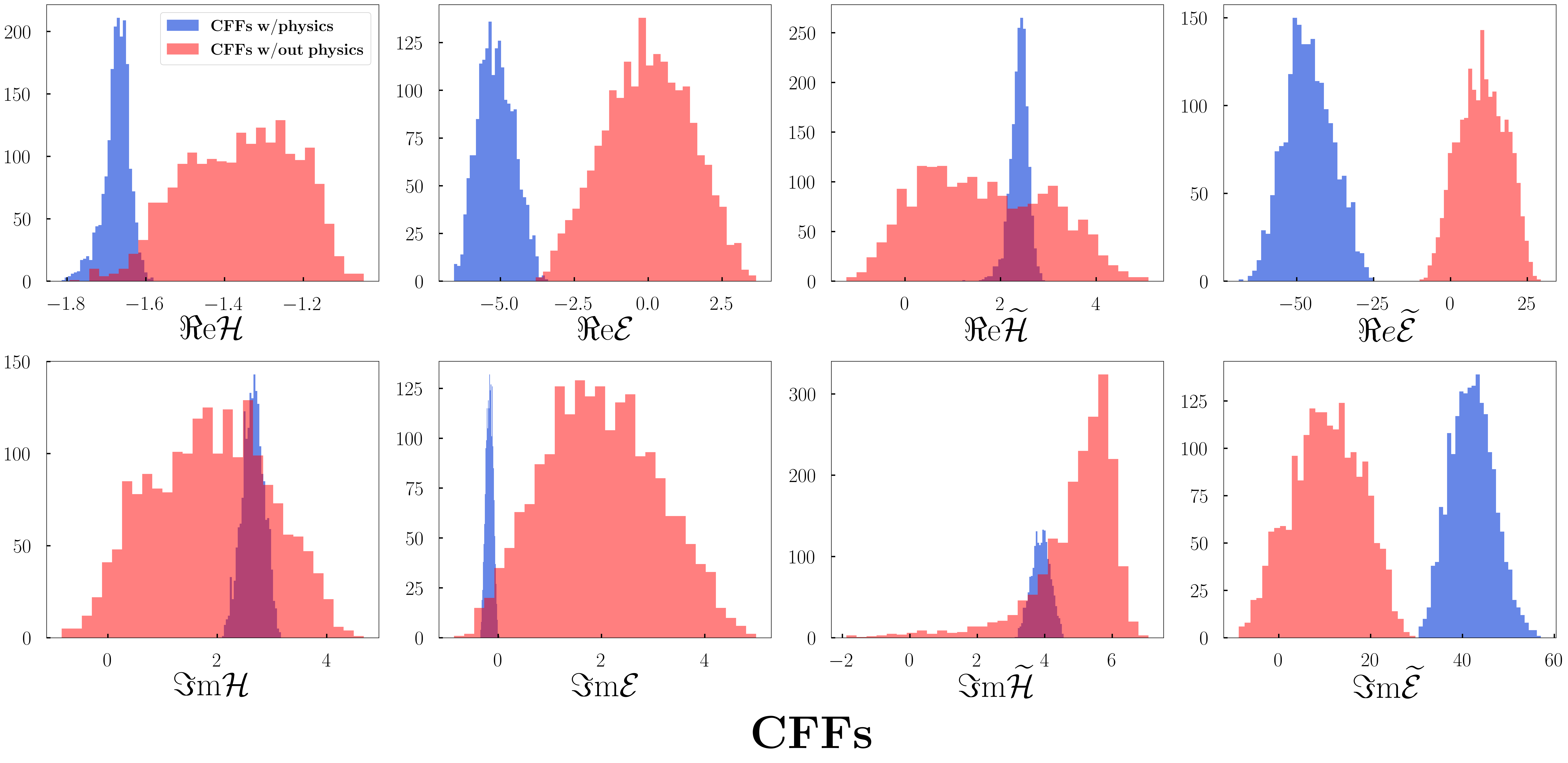}
\caption{Extracted CFFs using physics constrained input ranges calculated from a phenomenological model \cite{Kriesten:2021sqc,Goldstein:2010gu}   compared to uniformly distributed input ranges at  kinematics $x_{Bj} = 0.343$, $t=-0.172$ GeV$^{2}$, $Q^{2} = 1.820$ GeV$^{2}$.  }
\label{fig:ex1_physics}
\end{figure}

Finally in Fig.~\ref{fig:ex1_physics} we demonstrate a comparison of extracted CFFs from physics constrained inputs, versus uniformly distributed inputs. The uniformly distributed inputs with no physics constraints are the same inputs as in Fig.~\ref{fig:PredCffs}. To generate the physics constrained input CFF distributions, we utilized the spectator model-based parametrization of Refs.~\cite{Kriesten:2021sqc,Goldstein:2010gu}. 
The parametric forms for the various GPDs are fitted to PDF extractions in the forward limit as well as flavor separated form factor measurements in off-forward kinematics~\cite{Goldstein:2014aja}. 
Using the central values as the central GPD and smearing those values using a Gaussian over the standard deviation given by the uncertainty on the parameters, we can generate input GPD bands.
This GPD band for each GPD then contains all of the physics information from the fit in it. Propagating this physics information into our analysis, we calculate the CFFs from this GPD band and use those regions as inputs to the C-VAIM. 
One can see that the distributions with physics input are more tightly constrained, meaning there are smaller predicted widths. Quite a few of the distributions also do not overlap, this can be explained however because the spectator model is not constrained to any experimental data from DVCS measurements; therefore, it is unexpected to perfectly line up with the fits from data. 
One can notice that in the case of $\Re e \widetilde{\mathcal{E}}$ the sign of the CFF is contested between the 2 predictions. 

The latent $z$, or the output of the encoder model, is the key component of the C-VAIM as it attempts to learn the lost information in the forward mapping and converts the ill-posed inverse problem into a well-posed regression problem in the backward mapping. 
We analyze the latent space and its association with the CFFs using dimensionality reduction techniques. This allows us to explore the learned patterns which reflects the correlation between physics quantities, as well helps us to further understand the underlying physics.

We first utilize the principle component analysis (PCA), a linear dimensionality reduction algorithm. PCA works by finding the eigenvectors and eigenvalues of the covariance matrix of the feature space. By sorting the eigenvalues (and associated eigenvectors) of the covariance matrix from largest to smallest and truncating at a fixed input number of features, you can construct a projection matrix with which you can transform your input feature space. The benefits of PCA is that it is not only less computationally expensive for large datasets, it also preserve global structures in the dataset unlike other unsupervised dimensionality reduction techniques like t-distributed stochastic neighbor embedding (t-SNE). 

\begin{figure*}[ht]
\includegraphics[width=\linewidth]{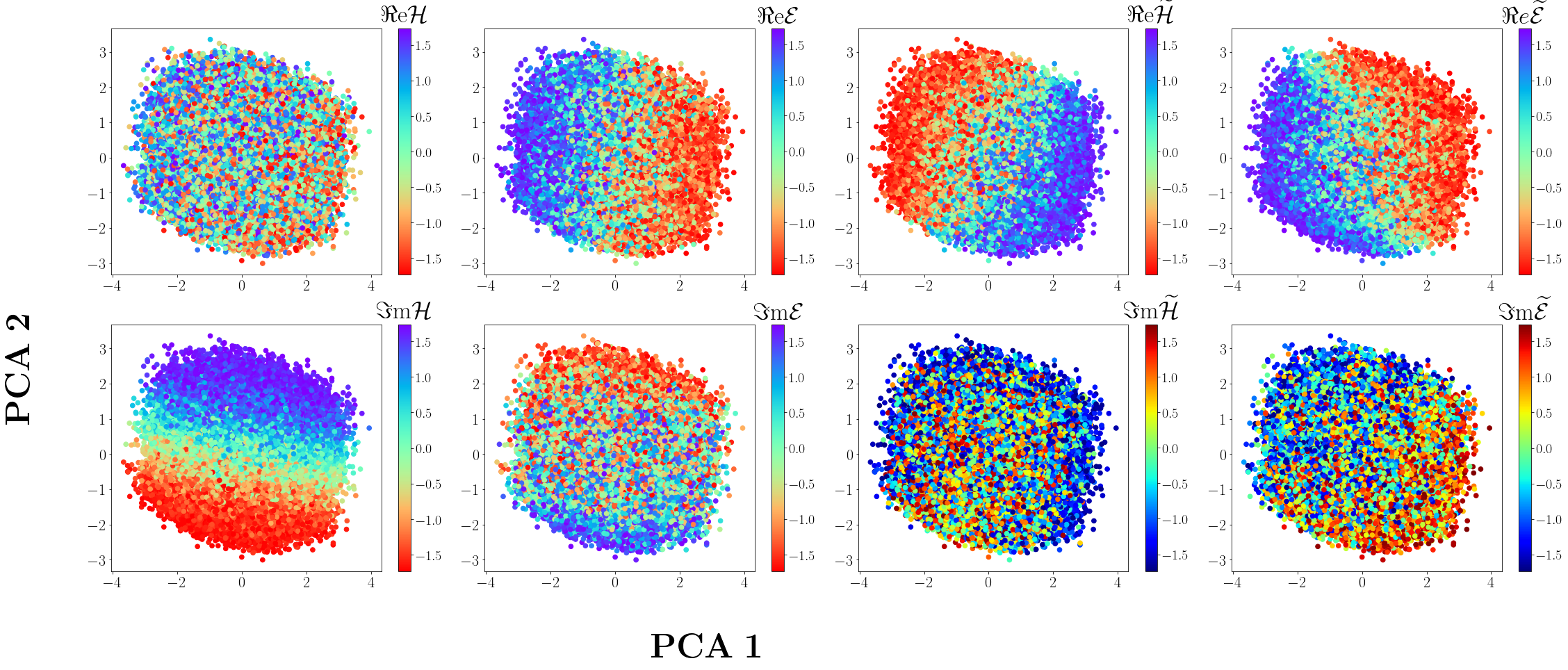}
\caption{The distribution of the 8 CFFs on latent space projected on the first two principle components for predicted CFFs at a kinematic value of $x_{Bj} = 0.343$, $t = -0.172$ GeV$^{2}$, $Q^{2} = 1.82$ GeV$^{2}$, and $E_{b} = 5.75$ GeV. 
}
\label{fig:latent_pca}
\end{figure*}

After our C-VAIM is trained, we project the latent variables onto the first two principle components. The distribution of CFFs samples is visualized on the latent space in fig.~\ref{fig:latent_pca}  for the first two PCA components (PCA 1, PCA 2). The color is the value of the CFF at the coordinates indicated by the specific PCA components. We can see that there is an emergence of some organization/pattern with respect to some of CFFs: specifically, $\Re e \mathcal{E}$, $\Re e \widetilde{\cal H}$, $\Im m {\cal E}$ and $\Im m {\cal H}$. This may indicate the latent space is learning to organize itself in such a way to encode specific information on those CFFs.

Taking, for instance, the CFF $\Im m \mathcal{H}$, the organization of that CFF into positive and negative values could indicate that in the mapping from CFF to cross section, what is lost is the sign information. Indicating that this CFF most likely appears to have a more significant impact on the CFF$^{2}$ term, appearing in the pure DVCS contribution, Eq.\eqref{eq:DVCSxsec}. 
A similar situation is seen for the CFFs $\Re e \mathcal{E}$ and $\Re e \widetilde{\mathcal{H}}$. 

It should also be noted that there is an obvious lack of structure in the latent space dimensional analysis for a few of the CFFs. 
In particular, we can look at a few specific cases of CFFs, namely $\Re e \mathcal{H}$, $\Im m \widetilde{\mathcal{H}}$, and $\Im m \widetilde{\mathcal{E}}$. 
The lack of structure in these cases could indicate separate issues that manifest in the same way in the latent space. 
In the case of $\Re e \mathcal{H}$ and $\Im m \mathcal{H}$, the randomness of the latent space distribution indicates that there is no lost information in the forward mapping from CFF to cross section, meaning that the cross section can exactly determine the value of those particular CFFs. 
This is corroborated by the results in Fig.\ref{fig:PredCffs} where both $\Re e \mathcal{H}$ and $\Im m \mathcal{H}$ have a small uncertainty width as compared to the input distribution. 
On the contrary, the behavior of the latent space dimensionality reduction analysis of $\Im m \widetilde{\mathcal{E}}$, where, from the theoretical knowledge of the exactly calculable, numerically small kinematic coefficient of this CFF, Eqs.~\eqref{eq:Ixsec} and \eqref{eq:DVCSxsec},  we can deduce that there is no information that can be extracted about it from the cross section. A similar situation appears for $\Re e \widetilde{\cal E}$, where, even if a clear pattern emerges, this cannot be readily interpreted as {\it e.g.} the emergence of ``sign information". Finally, $\Im m \widetilde{\mathcal{H}}$, displays a circular/elliptical pattern which can be interpreted as the emergence of either a sign change, or a change in slope of the CFF in $t$. This observation correlates well with the behavior displayed in Fig.\ref{fig:cff_trends1}, where $\Im m \widetilde{\mathcal{H}}$ is the only CFF with a slope which is sensibly different from zero.  

Our initial analysis suggests that gleaning physics information from the latent space is a very useful feature of the VAIM. 
Encouraged by these unexpectedly revealing results on the CFF analysis, in a future extension of the present analysis we will pursue a more refined approach including quantitative analysis methods for interpreting the latent space. 

\section{Conclusions}
\label{sec:conclusion}
The extraction of CFFs from DVCS cross section measurements has been performed by many groups using   various 
different methods  including artificial neural networks~\cite{Cuic:2020iwt,Moutarde:2019tqa,Grocholski:2019pqj,Kumericki:2019mgk} (for a review see Ref.~\cite{Kumericki:2016ehc}) and for a more recent extraction Ref.~\cite{JeffersonLabHallA:2022pnx}). In the broader scope of analyzing a large number of deeply virtual exclusive reactions planned at both Jefferson Lab and at the upcoming EIC, it is critical to understand exactly what information is contained in the cross section to provide guidance on which measurements would have the highest impact on global analyses. 
We studied unpolarized DVCS as a benchmark process that has been serving as a template in various analyses, and we reformulated the extraction of CFFs from the DVCS cross section as an inverse problem. Utilizing the VAIM architecture framework first developed in Ref.~\cite{Almaeen:2022ifg}, we investigated possible multiple solutions. Using a neural network to interpolate the cross section between kinematic points, we have also demonstrated that one can use the conditional VAIM (C-VAIM) architecture to make predictions of CFFs across ranges of kinematic values. 
Our predictions for the CFFs are summarized in Fig.~\ref{fig:cff_trends1}. The predictions are performed with a choice of prior described in Sec.~\ref{sec:VAIMCFF_procedure} covering the Jlab kinematic range of  $ 0.17 < -t < 0.37$ GeV$^2$, $0.34 <x_{Bj} < 0.38$, and $1.8 < Q^2 < 2 $ GeV$^2$ with individual bins listed in Table~I. 
Training with this limited data set was motivated by our need to focus on various sources of uncertainty of both epistemic and statistical nature, thus eliminating, at this stage, any possible systematics 
arising from the combination of different experiments. 

Our study includes a quantitative analysis of the impact of different models of the cross section~\cite{Belitsky:2010jw,Kriesten:2019jep,Kriesten:2020apm} on the CFF determination, as well as the implementation of the uncertainty quantification technique developed in Refs.~\cite{Grigsby:2020auv, Almaeen:2022imx} called random targets. 
We show that a large portion of the random target method uncertainty can be attributed to the reducible ML model uncertainty. Finally we used a phenomenological model to constrain the input CFFs to the VAIM architecture and simultaneously study the effects of physics inputs on the predicted CFFs. 
Our extracted CFFs and uncertainty bands are compared to the ANN-based ones in Ref.~\cite{Cuic:2020iwt}. 
A crucial component of this inverse-problem solver methodology is the dimensionality reduction analysis of the latent space. Through PCA on the latent variables we can interpret - in a human readable way - what information is lost in the forward mapping from CFF to cross section. 
Unlike standard fitting methods where we lose this crucial physics knowledge, using the VAIM we can try and decode this information to give us a deeper understanding of the physical observables. 

Future developments will include extending the reach of our physics aware neural networks by using {\it e.g.} lattice QCD results on both the moments of GPDs and on their $x$ dependence, according to the approach developed in Ref.~\cite{Almaeen2021}, as well as introducing additional methods to address uncertainty quantification.
Finally, we will further develop the methods to access information from the latent space specific for the observables of DVCS-type processes, thus opening a brand new way of quantifying and interpreting these, so far, elusive quantities. 
A more in depth treatment of the information available in data will provide guidance on where in the kinematics phase space 
should measurements of deeply virtual exclusive experiments be placed to make high-impact, pivotal contributions to the GPD global analysis.

\acknowledgements
This work was funded by DOE grants DE-SC0016286 (D.A., S.L.), PHY 1653405,  PHY 2209424 and  Research  Corporation  for  Science  Advancement through the Cottrell Scholar Award (H.L.), the  
SURA Center for Nuclear Femtography (B.K., S.L., Y.L.) and a PhD scholarship from Jouf University, Saudi Arabia (M.A.). T.A. was supported by a Ph.D. scholarship from Al-Baha University, Saudi Arabia. The work of B.K. at Argonne National Laboratory was supported by the U.S.~Department of Energy under contract DE-AC02-06CH11357.

\appendix
\section{Compton Form Factors}
\label{sec:app_cff}
At leading order the perturbatively calculable Wilson coefficient functions read:

\begin{eqnarray}
\mathcal{F}^{q}(\xi,t) &=& \mathcal{C}^{+}(x,\xi;Q^{2}) \otimes F^{q}(x,\xi,t) \nonumber \\
&=&   e_{q}^{2}\int_{-1}^{+1}dx \Big[\frac{1}{\xi - x - i \epsilon} - \frac{1}{\xi + x - i\epsilon} \Big] F^{q}(x,\xi,t) \nonumber \\
\label{eq:cff1}
\end{eqnarray}
 and similarly with the polarized Compton form factors
\begin{eqnarray}
\widetilde{\mathcal{F}}^{q}(\xi,t) &=& \mathcal{C}^{-}(x,\xi;Q^{2}) \otimes \widetilde{F}^{q}(x,\xi,t) \nonumber \\
&=&   e_{q}^{2}\int_{-1}^{+1}dx \Big[\frac{1}{\xi - x - i \epsilon} + \frac{1}{\xi + x - i\epsilon} \Big] \widetilde{F}^{q}(x,\xi,t) \nonumber \\
\label{eq:cff2}
\end{eqnarray}
where GPDs $F \in \{H,E\}$.

One can separate out the real and imaginary parts of the Compton form factor 

\begin{eqnarray}
\Re e \mathcal{F}(\xi,t) &=& e_{q}^{2}  \,P.V. \int_{-1}^{+1} dx \Big[\frac{1}{\xi - x} - \frac{1}{\xi + x } \Big] F^{q}(x,\xi,t) \nonumber \\ \\
\Im m \mathcal{F}(\xi,t) &=& \pi e_{q}^{2} \Big( F^{q}(\xi,\xi,t) - F^{q}(-\xi,\xi,t)  \Big)
\end{eqnarray}
where P.V. is the principal value integral. 

\section{Multi-Layered Perceptrons}
\label{appendixB}
MLPs are a series of fully connected linear layers with non-linear activation functions that are applied at each layer. The \textit{parameters} of the $j$th layer are the matrix of network weights $W_j$ and bias vector $b_j$. The weights can be thought of as the strength or importance of each node during learning while the bias allows for a more robust neural connection during learning. Layer $j$ projects its input $x_{j}$ to $x_{j+1}$  with a linear transformation followed by a non-linear activation function $z_j$ as:
\begin{eqnarray}
x_{j+1} = z_j(W_{j}x_{j} + b_j)
\end{eqnarray}

\begin{figure}
    \centering
    \includegraphics[width=8.5cm]{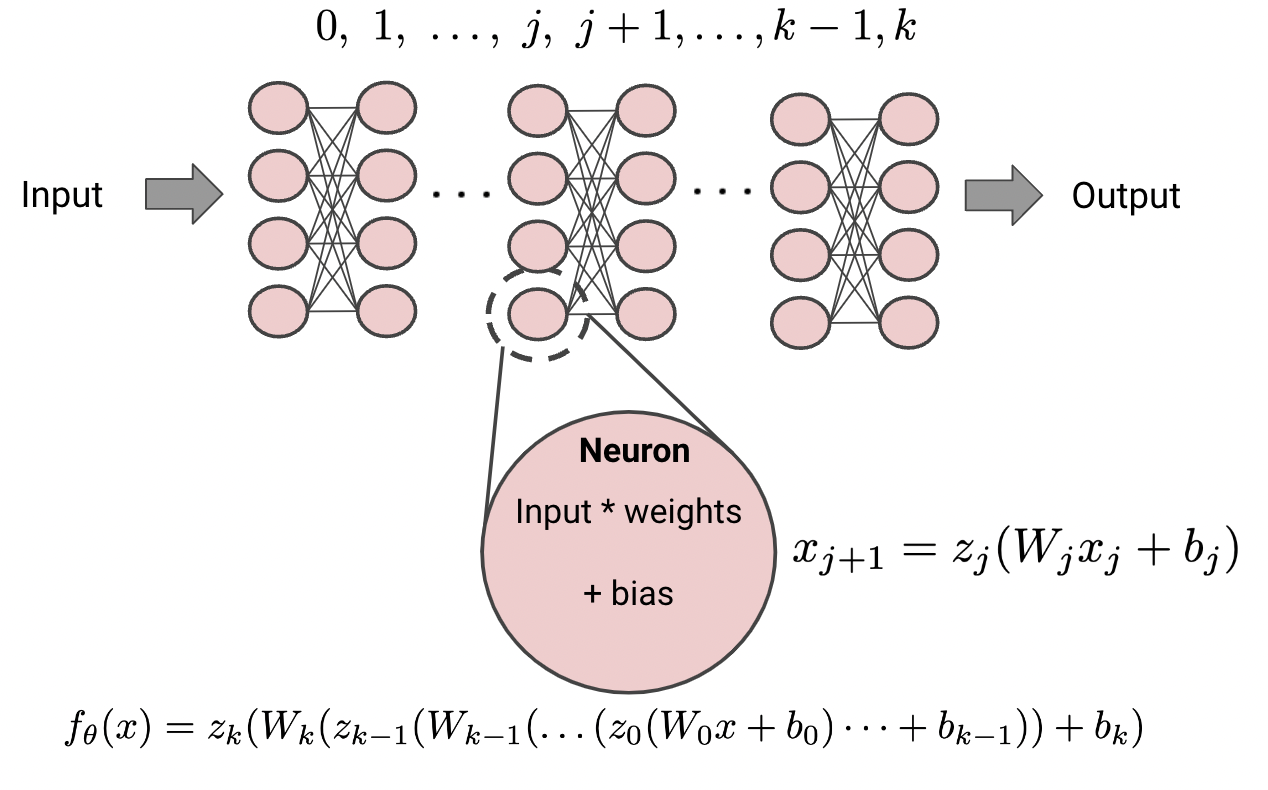}
    \caption{Architecture of a general DNN MLP in which each layer is fully connected series of neurons. Each neuron consists of an input $x_{j}$, a weight $W_{j}$, and a bias $b_{j}$. The output of the neuron $x_{j+1}$ that connects to the next series of nodes is passed through a non-linear activation function $z_{j}$.}
    \label{fig:dnn_general}
\end{figure}

A common choice of activation function used in the literature is the rectified linear unit (ReLU) 
\cite{DBLP:journals/corr/abs-1803-08375}
defined as $z_j(x) := \text{max}(x, 0)$ and its variants such as Leaky RELU~\cite{Maas2013RectifierNI}. This is due to solving a common training problem known as the vanishing Gradient problem during backpropagation and gradient update as contrasted to other activation functions such as the hyperbolic tangent. The non-linear activation function is what allows the neural network to learn complex representations of the mapping from inputs to outputs. Without the non-linear activation function the MLP can be reduced to a singular linear layer and is not a useful functional approximation method. The full MLP, $f$, can be written as a composition of $k$ layers:

\begin{eqnarray}
f_{\theta}(x) &=& z_k(W_{k}(z_{k-1}(W_{k-1}(\dots (z_0(W_{0}x+b_{0})\dots\nonumber \\
&&+b_{k-1}))+b_{k}) 
\end{eqnarray}
\noindent where $\theta$ denotes the set of all parameters $\{(W_j, b_j) \}$ that make up the deep neural network (DNN). Ideally, deep networks allow for each layer learn a slightly improved representation of the input data, until accurate predictions are made in the final linear layer $k$. The choice of activation functions $z$, the number of layers $k$, and the size of each $W$ and $b$ parameter are known as hyper-parameters, the set of which fully defines a network ``architecture." The hyper-parameters can be tuned using a hyper-parameter search algorithm such as the hyperband algorithm \cite{DBLP:journals/corr/LiJDRT16}. We demonstrate a simple DNN in Figure \ref{fig:dnn_general} where we show the level of abstraction at each fully connected layer including the purpose of the nodes.

DNNs are trained to satisfy some metric of performance given by a \textit{loss function}, $\mathcal{L}$, that represents the accuracy of their predictions. One common loss function in many neural networks, especially for regression problems, is the mean squared error (MSE). During each epoch of learning, given a dataset of $N$ pairs of inputs and predictions $\{x^{(i)}, y^{(i)}\}_{i=0}^{N}$, the loss function is computed in batches of $B$ samples using our current parameters $\theta_t$. During the step of backpropagation, we differentiate the loss function with respect to $\theta_t$ and use the gradient vector to shift our parameters in a direction that has lower prediction error known as stochastic gradient descent (SGD): 
\begin{eqnarray}
\label{sgd}
    \mathcal{L}_{\theta_t} &=& \frac{1}{2}(y - f_{\theta_t}(x))^2 \\
    \theta_{t+1} &=& \theta_t - \alpha \nabla_{\theta_t}\frac{1}{B}\sum_{i=0}^{B}\mathcal{L}_{\theta_t}(x^{(i)}, y^{(i)}) 
\end{eqnarray}
\noindent SGD is controlled by a parameter $\alpha$ which is a small scalar known as the \textit{learning rate}. In practice, it is common to treat $\alpha$ as a vector of size $|\theta_t|$ and adjust each element with various heuristics such as momentum across multiple updates \cite{kingma2014adam}. 

DNNs can have millions of parameters and therefore are at risk of over-optimizing on training inputs, this comes at the expense of their ability to generalize to unseen data. This problem can be addressed in a variety of ways; we control this problem through \textit{regularizing} the network during training. We choose a regularization technique called dropout \cite{srivastava2014dropout}, where elements of a layer's weight matrix $W$ are zeroed with probability $p$ by a binary mask that is randomly generated before each training step. Dropout prevents a network from becoming overly reliant on the connections between specific neurons. Dropout also allows us to estimate the systematic error of the neural network \cite{gal2015dropout}, where the distribution of a number of predictions with different dropout masks can approximate variational inference. We periodically evaluate the performance of a subset of \textit{training} data against a \textit{validation} subset that the network has never seen during the training procedure. This allows us to monitor network generalization capabilities and simultaneously stop the learning procedure before over-fitting occurs.  Another approach reduces the network's ability to overfit to the training data by constraining the parameters to be close to the origin; this can be done by adding the L2 norm $||\theta_t||_2$ to the loss function (Eq. \eqref{sgd}) with coefficient $\lambda$.

The MLPs we discussed above are supervised network algorithms in which the inputs and outputs of the network are known and can be compared to data; however, our method not only relies on DNN MLPs but also an unsupervised method of reconstructing lost information in a \textit{latent space}. This unsupervised network architecture is known as a variational autoencoder (VAE) a subset of a more general class of generative unsupervised algorithms called autoencoders (AE). Since these networks are not as well known in the literature of nuclear phenomenology we will give a brief introduction to them here.


\section{Autoencoders}
\label{appendixC}

\begin{figure}
    \centering
    \includegraphics[width = \linewidth]{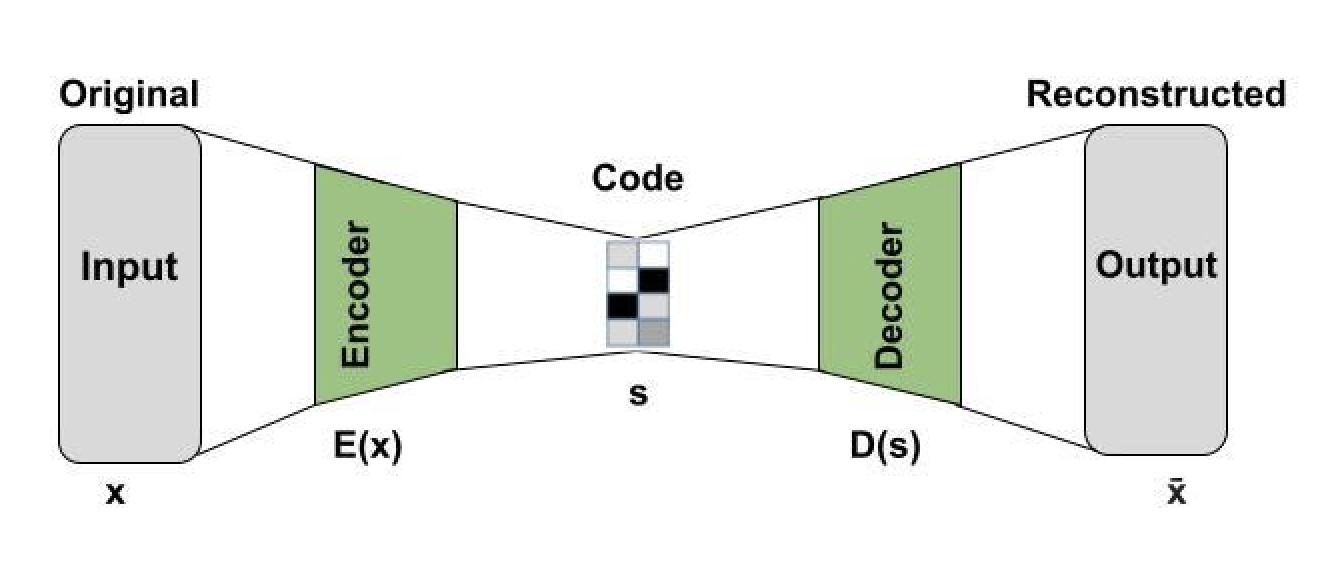}
    \caption{General architecture of the autoencoder}
    \label{fig:AEs}
\end{figure}

An autoencoder is a form of feedforward neural network that is primarily aimed to encode the input into a compact and meaningful representation and then decode it so that the reconstructed input is as close to the original as possible \cite{rifai2011contractive}. It is used in unsupervised learning to learn efficient data coding. Data approximations, encoding, and data representations are the primary goal of the Autoencoders. 
To build an Autoencoder, there are three necessary components (Figure \ref{fig:AEs} ). The first component is known as the "encoder", which is responsible for determining the appropriate encoding (latent space representation), which is the second component of the Autoencoder, from the input data in a compressed form without losing as much important information from the original data as possible. The third component of the AE is learning how to reproduce the original data, and it is known as the "decoder", which uses the encoding layer to provide an output of the Autoencoder that is comparable to the input data.
 We define the encoder mapping function as $E:x \rightarrow s$, where  x and the s represent the input and latent representation, respectively. As a result, the decoder's mapping function is demonstrated by $D:s \rightarrow \hat{x}$. 

Given a training set of data $ t=\{x_{1},x_{2},x_{3},... x_{N}\}$, each sample is represented in the latent-space representation (coding layer) as $s_{n}= E( x_{n}),n \in \{1,2,3,…,N\}$, however, the reconstructed sample is provided by $ \hat{x}_{n}=D( s_{n} )=D(E( x_{n} )),n \in \{1,2,3,..,N\}$. 

Minimizing the error (such as Mean Square Error (MSE)) between original samples and reconstructed ones is the goal of the autoencoder, which is analogous to the reconstruction error:
\begin{eqnarray}
\sum_{n=1}^{N}  || \hat{x}_{n} - x_{n} ||^{2}
\end{eqnarray}
Representation learning through autoencoders has been used for a variety of applications such as classification, clustering, anomaly detection, recommendation systems, and dimensionality reduction.
One of the autoencoder's shortcomings in image applications, for example, is that the quality of the output results is generally blurry. The use of the loss function that calculates the difference between the original data and the reconstructed data, is the main reason since it doesn't take the prior knowledge into consideration.
Another issue is that there are no constraints on the latent space's values or distribution (i.e., latent space is not regularized), and sampling from areas of the latent space that do not correlate to any data point would yield nonsensical results.
The decoder, in general, cannot be utilized as a generative model for the entire space due to the non-regularized latent space. Therefore, the primary use of an autoencoder is compression.

\bibliography{BIB}
\end{document}